\documentclass[12pt,preprint]{aastex}
\begin{document}

\title{Beryllium Abundances in Stars of One-Solar-Mass}

\author{Ann Merchant Boesgaard\altaffilmark{1}}
\affil{Institute for Astronomy, University of Hawai`i at M\-anoa, \\
2680 Woodlawn Drive, Honolulu, HI{\ \ } 96822 \\ }
\email{boes@ifa.hawaii.edu}

\author{Julie A. Krugler\altaffilmark{1}}
\affil{Department of Physics \& Astronomy, \\
Michigan State University, East Lansing, MI 48824-1116} 
\email{kruglerj@msu.edu}

\altaffiltext{1}{Visiting Astronomer, W.~M.~Keck Observatory jointly operated
by the California Institute of Technology and the University of California.}

\begin{abstract}
We have determined Be abundances in 50 F and G dwarfs in the mass range of 0.9
$\leq$ M$_\odot$ $\leq$ 1.1 as determined by Lambert \& Reddy.  The effective
temperatures are 5600 to 6400 K and metallicities from $-$0.65 to +0.11.  The
spectra were taken primarily with Keck I + HIRES.  The Be abundances were
found via spectral synthesis of Be II lines near 3130 \AA.  The Be abundances
were investigated as a function of age, temperature, metallicity and Li
abundance in this narrow mass range.  Even though our stars are similar in
mass, they show a range in Be abundances of a factor of $>$40.  We find that
[Be/Fe] has no dependence on temperature, but does show a spread of a factor
of 6 at a given temperature.  The reality of the spread is shown by two
identical stars which differ from each other by a factor of two only in their
abundances of Li and Be.  Our thin-disk-star sample fits the trend between Be
abundance and [Fe/H] found for halo and thick disk stars, extending it to
about 4 orders of magnitude in the two logarithmic quantities.  Both Fe and Be
appear to increase similarly over time in the Galaxy.  One-third of our sample
may be classified as subgiants; these more-evolved stars have lower Be
abundances than the dwarfs.  They have undergone Be depletion by slow mixing
on the main sequence and Be dilution during their trip toward the red giant
base.  There are both Li and Be detections in 60 field stars in the
``Li-plateau'' of 5900 - 6300 K now and the abundances of the two light
elements are correlated with a slope of 0.34 $\pm$0.05, with greater Li
depletion than Be depletion.
\end{abstract}


\section{INTRODUCTION}
The study of the light elements reveals the physics of a star beneath its
surface.  Lithium, Be, and B can all be used as probes to understand surface
mixing down to temperatures of about 2.5 $\times 10^{6}$, 3.5 $\times 10^{6}$,
and 5 $\times 10^{6}$ K respectively, where they are destroyed by fusion
primarily with protons.  Boron provides the deepest look into
the interior of a star, but the B I, B II, and B III resonance lines appear
only in the ``satellite'' UV so it is impossible to find B abundances without
going above the atmosphere.  Lithium is markedly easier to observe given that
its resonance doublet occurs in the red portion of the spectrum at 6708 \AA{},
but inasmuch as there are multiple sites for the creation and destruction of
Li, the interpretation of the Li abundance can be complicated.  Beryllium
provides a clearer picture as it has one mechanism of production: spallation
reactions in the vicinity of supernovae or in the ambient interstellar gas.
The resonance lines of Be II are found at 3130.421 and 3131.065 \AA{}, which
are still observable with ground-based telescopes.  However, the Be II
features are near the atmospheric cutoff ($\sim$3000 \AA{}), so stars must be
observed at low air mass.

There have been several recent studies of the light elements Li and Be.
Boesgaard et al.~(2004a) (hereinafter B04a) investigated the correlation of Li
and Be in F and G dwarf stars in cluster and field stars in the metallicity
range of $-$0.45 $\leq$ [Fe/H] $\leq$ +0.11.  Included in the paper was a plot
of the Be abundance against T$_{\rm eff}$ from a variety of cluster and field
sources.  The plot showed a large spread in A(Be) = log N(Be)/N(H) + 12.00 and
the focus of their study was narrowed by defining two temperature regimes: the
cool side of the Li-Be dip (Boesgaard \& Tripicco 1986; Boesgaard \& King
2002) from 6300-6650 K (where Li increases with decreasing temperature) and
near the Li plateau from 5900-6300 K (where Li begins to decline with
decreasing temperature).  B04a found that Li was depleted more than Be in both
temperature regimes.  The amount of Li and Be depletion in the Li-Be dip
(T$_{\rm eff}$ = 6300 - 6650 K) is attributed to slow mixing and is dependent
upon age and temperature.  Older stars showed more Li and Be depletion and the
hotter stars, near the bottom of the dip, show the greatest depletions.  There
was no difference in the depletion patterns between the cluster and field
stars.

Santos et al.~(2004a) (hereafter S04a) determined Be abundances by spectrum
synthesis in field stars ranging from 4800-6300 K.  They found the Be
abundance to peak at $\sim$6100 K, with decreasing Be abundances both towards
the cooler stars and towards the Li-Be gap that encompasses hotter stars.
They suggest that the peak may be explained by a the bias in their sample of
metal-rich stars which would also have enhanced Be due to Galactic enrichment.
They discuss the decline in the Be abundance and the possibility of a Be
plateau around T$_{\rm eff}$ $\sim$5600 K.  This decline and plateau are
compared to the Be depletion models that introduce rotational mixing by
Pinsonneault et al.~(1990) for 1.7 Gyr ages.  The match is acceptable above
5600 K, but there is a better fit for cooler temperatures with the wave-driven
mixing models of 4.5 Gyr of Montalban \& Schatzman (2000).  This same data set
was also analyzed in Santos et al.~(2004b) to find a possible correlation
between Be abundance and planet hosts.  It was found that planet hosts have a
``normal'' Be abundance.

Lambert \& Reddy (2004; hereafter LR04) have compiled a database of Li
abundances for 451 F and G thin disc stars.  The stars were taken from three
different surveys of Li in field stars: Reddy et al.~(2003), Chen et
al.~(2001), and Balachandran (1990).  The stars are located in the solar
neighborhood and have measured Hipparcos parallaxes.  HR diagrams were created
for these field stars using the parallaxes to calculate M$_{\rm V}$.  The
majority of these stars were slightly evolved away from the zero age main
sequence, so it was possible to calculate evolutionary ages from the Girardi
et al.~(2000) (hereafter G00) isochrones.  The corresponding evolutionary
tracks were used to determine masses for each metallicity subset.

It has been shown to be useful to study more than one of the light elements at
once to develop a fuller picture of the stellar mixing and depletion
processes.  For example, B04a investigate the correlation of Li and Be, and
Boesgaard et al.~(2005) discovered the correlation of Be and B.

The location of the Li resonance lines in the red spectral region makes Li
easier to observe and measure than Be, due to the location of the Be II lines
near the atmospheric cut off and due to the substantial blending found in the
UV in solar-type stars.  Given that the LR04 data have a complete set of
stellar parameters, along with Li abundances, they provide a good base from
which to study the Be abundances.  In order to disentangle information
regarding the Be and Li abundances from the large spread in the B04a plots of
A(Be) vs.~T$_{\rm eff}$ and A(Be) vs.~[Fe/H], we placed a one solar mass
constraint on the LR04 data in our star selection.  In this way we can
understand better the nature of the processes which affect both Li and Be at 1
M$_\odot$.

\section{OBSERVATIONS}
A total of 52 high resolution spectra of the Be II resonance lines at 3130.421
and 3131.065 \AA{} were taken over several nights with the HIRES instrument on
the Keck I 10m telescope and the coud\`e spectrograph on the
Canada-France-Hawaii 3.6m telescope (CFHT).  These telescopes are on Mauna Kea
at 14,000 feet elevation which is above $\sim$40\% of the Earth's atmosphere.
This allows additional throughput in the ultraviolet region compared to other
observatory sites.  Additionally, these spectra were taken when the stars were
near the meridian, reducing the air mass and reducing the effects of
atmospheric dispersion.

\subsection{Star Selection}

The stars in this study were chosen from the LR04 paper in which 451 F and G
dwarfs were analyzed for the purpose of understanding the metallicity, mass,
and age effects on the Li abundance.  Their study determined ages and masses
using the G00 tracks, and Hipparcos parallaxes for M$_{\rm V}$.  Our study
focused on the stars that LR04 found to be within 0.9 $\leq$ M$_\odot$ $\leq$
1.1.  Of the 451 stars in the LR04 study there were 156, or 35\% in that mass
range; we have Be observations of one third of these stars.

Figure 1 is an HR Diagram showing the positions of our stars.  They are
separated into dwarfs and subgiants by the criterion of S04 at log g = 4.1.
The circled dot indicates the Sun.  They range from 5600 $<$ T$_{\rm eff}$ $<$
6600 and most are somewhat evolved away from the zero age main sequence.  Due
to the fact that most of the stars in LR04 are evolved beyond the zero age
main sequence, it was possible for them to determine stellar ages.

\subsection{New Keck/HIRES Observations} 

In 2004, the HIRES spectrometer was upgraded with a 6144 x 4096 mosaic CCD,
which is split into three chips with one optimized for the ultraviolet/blue
end of the spectrum.  The quantum efficiency is 94\% for this chip in the
beryllium region, about 12 times better than the old chip.  Between 31 January
2005 and 9 June 2007, 40 Keck spectra were taken with the new HIRES detector
on nine separate nights.  Given the relative brightness of these stars, they
were taken in evening and morning twilights and on partly cloudy nights.  The
decker used was B5 which has slit dimensions of 3.5 x 0.861 arcsec.  These
data have a spectral dispersion of 0.0132 \AA{} pixel$^{-1}$ and a measured
spectral resolution of $\sim$49,000 near 3130 \AA.  Exposure times ranged from
3-30 minutes to attain a signal-to-noise ratio (S/N) range from 33 to 277 and
a median S/N of 146 and a mean of 168.

On each of the nine nights on which observations were made for this program,
calibration frames were also taken.  Typically 7-9 bias frames were recorded
and Th-Ar spectra were obtained at the beginning and end of each night.  The
internal flat fields were taken with different exposure times for each of the
three CCDs.  In order to achieve sufficient flux for the Be II line region
exposures of 50 s were needed for the UV chip.  The optimum exposure for the
green chip was typically 3 s and that exposure would be useful for the longer
orders of the UV chip.  In order not to saturate the flat fields on the red
chip exposures of 1 s were obtained.  Seven to nine flat exposures were taken
at each of the three exposure times to achieve the proper flux for the various
orders on all three CCDs.

Table 1 gives the star name, V magnitude, date of observation, exposure time,
and total S/N of the stars for which observations were taken using Keck I with
the upgraded HIRES.  

\subsection{Other Observations}
The Keck spectra from the upgraded HIRES were supplemented by spectra in the
LR04 one-solar-mass range from the original HIRES and the CFHT.  With the
original HIRES on Keck I with a Tektronix 2048 x 2048 CCD (Vogt et al. 1992),
five stars were observed between 13 November 1999 and 5 January 2002.  These
data have a spectral resolution of 0.022 \AA{} pixel$^{-1}$ which gives a
resolution of $\sim$48,000 per pixel.  Exposure times for those stars ranged
from 5-12 minutes with a median S/N of 140 and a mean of 122.

Five stars were observed on 15 and 16 October 1995 on the 3.6m CFHT with the
Gecko f/4 spectrograph with a resolution of $\sim$120,000 and a dispersion of
0.010 \AA{} pixel$^{-1}$.  The detector was a 2048 x 2048 array with 15 $\mu$m
pixels that is UV sensitive with a quantum efficiency 80\% in the beryllium
region.  Exposure times for these five stars range from 15-45 minutes, with a
median and mean S/N of 72.  Information on these observations and data
reduction can be found in Boesgaard et al.~(2001).  One additional star (HD
142860) was obtained with the f/8 coud\'e spectrograph at CFHT at a resolution
of 24,000.  Details of that observation can be found in Boesgaard et
al.~(2004b).

Table 2 includes the CFHT stars and Keck I stars that were previously included
in Boesgaard et al.~(2001, 2004a, 2004b).  It lists the name, V magnitude,
date of observation, exposure time, total S/N, telescope, and the reference to
the paper from which the star was taken and where the earlier analyses were
presented.  The Be abundances for these stars have been redetermined here
using the parameters determined by LR04.

\subsection{Data Reduction}

The Keck I data from the upgraded HIRES were reduced using IRAF and using two
different reduction pipelines: the new MAKEE Data Reduction Software and the
IDL-based HIRES Redux software.  Both software routines were designed for the
new chip.  We used our own master flat fields (one for each chip), biases, and
arcs rather than adopt archival ones.  It is critical to have long flat field
exposures, typically 50 s,in order to have sufficient flux at the UV
wavelengths of the Be II lines, while 3 s is enough for the green chip and 1 s
for the red chip.  Our wavelength calibrations were found from our Th-Ar
exposures at the beginning and end of the each night.  The wavelength solution
was applied for the each night.  Cosmic ray removal was also performed on the
stars in the Be region using the IDL version of
L.A. Cosmic\footnote{http://www.astro.yale.edu/dokkum/lacosmic}.  The data
were then doppler corrected and continuum fit in IRAF using the onedspec
package and the echelle package.

The data reduction for the stars discussed in $\S$2.3 has been described in
the original papers; see Boesgaard et al.~2001, B04a, and B04b for the details
of the procedures used regarding the CFHT stars and the earlier Keck I HIRES
stars.

\section{ABUNDANCES}

\subsection{Stellar Parameters}
In order to measure the Be abundance in these stars, we interpolated
atmospheric models from the Kurucz grid point models (Kurucz 1993).  These
require metallicity ([Fe/H]), effective temperature (T$_{\rm eff}$), surface
gravity (log g), and microturbulence ($\xi$).  The [Fe/H], T$_{\rm eff}$, and
log g were adopted from LR04; however, the microturbulence parameter was not
included.  Microturbulences were determined using the following relation
derived by Edvardsson et al. (1993):

$\xi$ = 1.25 + 8 $\times$ 10$^{-4}$(T$_{\rm eff}$ $-$ 6000) $-$ 1.3(log g $-$ 4.5)

A complete list of the stellar parameters (T$_{\rm eff}$, log g, [Fe/H], and
$\xi$) for all our stars is found in Table 3.

\subsection{Beryllium Synthesis}
We determined the abundance of Be by fitting the Be II resonance lines at
3130.421 and 3131.065 \AA{} using the spectral synthesis software
MOOG\footnote{http://verdi.as.utexas.edu/moog.html} (Sneden 1973) as updated
in 2002 (in ``synth'' mode).  Due to the considerable blending around the
3130.421 \AA{} line the stronger of the doublet, the fit of the 3131.065 \AA{}
line gave a more reliably-determined Be abundance.

Our line list includes 307 atomic and molecular lines between 3129.5 and
3132.5 \AA.  It was adopted largely from Kurucz to best fit the Be region with
some modifications to the list from former studies of Be abundances in stars
covering a wide range of temperatures and gravities., e.g. the log $gf$ value
of the 3129.763 \AA{} Zr II line.  During the synthesis-fitting a Gaussian
profile was used for the instrumental and other broadening.  The instrumental
broadening has a full-width half-maximum of 0.08 \AA{} and most of the stars
were fit with FWHM 0.08 to 0.10 \AA\.  The largest broadening was 0.18 \AA{}
for HD 218470, an F5 V star in the Li-Be dip region with T$_{\rm eff}$ = 6495
K.

Figures 2-5 show examples of the synthetic spectra that were created for each
star.  For each spectrum, the dots represent the data points, while the solid
line is the synthesis with the best fit.  The dash-dotted lines represent a
factor of two in the Be abundance above and below the best-fit Be abundance
and the dotted line represents a synthetic spectrum without any Be.  The Be
abundance is represented in terms of A(Be); the meteoritic value is A(Be)=1.42
(Grevesse \& Sauval 1998).

In Figure 2 HD 693 and HD 103799 are compared.  They have essentially the same
T$_{\rm eff}$, [Fe/H], age (6.1 Gyr), and A(Be).  The main discernible
difference is in the rotation as is seen in the line broadening in HD 103799
where its full-width half-maximum (FWHM) gaussian is 0.13 compared to that of
HD 693 at 0.09 .  Shown in Figure 3 are HD 186379 and HD 94012 which have
nearly identical [Fe/H] and the same A(Be) and vary only slightly in age (10
Gyr vs.~8 Gyr) even though they differ in T$_{\rm eff}$ by 250 K.  In Figure 4
are HD 199960 and HD 15335.  They have the same age (6.5 Gyr), and T$_{\rm
eff}$; however, they differ in both [Fe/H] and A(Be) by a factor of two.  This
supports the trend of an increasing Be abundance with increasing [Fe/H] for
halo and disk stars (e.g. see Boesgaard \& Novicki 2006 Figure 7).  Finally,
in Figure 5 are HD 209320 and HD 30743.  These stars have the same age but
differ in [Fe/H] by 0.4 dex. T$_{\rm eff}$ by 225 K, and in A(Be) by 0.2 dex.

Table 4 includes the parameters such as mass, age and absolute visual
magnitude as determined by LR04, including their value for A(Li).  The final
column has the Be abundance determined in this work.

In the temperature regime of our stars, an error of 100 K results in a
negligible error in A(Be), $\leq$ 0.01 dex.  However, an uncertainty in log g
of $\pm$ 0.5 produces an error between 0.044 to 0.052 dex depending upon
T$_{\rm eff}$.  A combination of errors due to uncertainties in T$_{\rm eff}$,
log g, [Fe/H], $\xi$, S/N of the observed spectrum, and the goodness-of-fit of
the synthetic spectrum combine to give errors in A(Be) of 0.1 to 0.2 dex.
There are nine stars analyzed here with Be detections that can be compared
with the results in B01, B04a, and B04b.  The average differences in the
parameters used are $\Delta$T = 5 K, $\Delta$log g = $-$0.05, $\Delta$[Fe/H] =
$-$0.03 (in the sense of LR04 minus previous publications); these are
insignificant differences.  The differences in the values of A(Be) for the
nine stars are $-$0.12 $\pm$0.12.

\section{ANALYSIS AND RESULTS}

\subsection{General Results}

We have found Be abundances in 46 of our 50 stars ranging from A(Be) = 0.57 up
to the meteoritic abundance of 1.42.  This represents a spread in A(Be) of
almost an order of magnitude in one solar-mass stars.  The four upper limits
to A(Be) extend this range down to $<$$-$0.2 or a factor of more than 40.  In
our sample there are six stars with only upper limits on the Li abundance and
four of them also have only upper limits on Be, while the two others have
measured Be.  Two of the Li- and Be-deficient stars, HD 182101 and HD 218470,
are in the Li-Be dip region of T$_{\rm eff}$.

Histograms of the Li and Be abundance distributions are shown in Figure 6.
For this sample of 44 Li detections of one solar-mass stars the median value
of A(Li) = 2.39 and for the 46 Be detections the median A(Be) is 0.92.  Both
of these values are lower than their meteoritic counterparts of 3.31 and 1.42
respectively; this indicates considerable stellar depletion relative to the
meteoritic values.  All the stars are Li depleted and all but one are Be
depleted.

When we narrow the mass range to those stars that are within $\pm$0.03 solar
masses (0.97 - 1.03 $M_{\odot}$) and select those within $\pm$100 K of the
Sun, there are six in our sample.  For these six the mean A(Li) = 2.24
$\pm$0.23 and the mean A(Be) = 1.03 $\pm$0.14.  These values compare well,
within the errors, with the medians from the whole sample.  The six stars (HD
22521, 26421, 94835, 100180, 186379, and 217877) are all older than the sun by
3-4 Gyr (except HD 94835 which is only 2.3 Gyr older) according to the G00
tracks and isochrones.  Their old age and relatively high mean Li abundance,
A(Li) = 2.24, only deepens the mystery of the large solar Li depletion,
A(Li)$_{\odot}$ $\sim$1.0.

\subsection{HR Diagram}

In the LR04 study, masses and ages were determined via the G00 isochrones and
evolutionary tracks.  We have put our Be results in the familiar context of the
HR diagram by showing the relative Be abundances along with the G00
evolutionary tracks and isochrones in Figures 7 and 8.  We have used tracks
for [Fe/H] of 0.0 and $-$0.4 and include stars with [Fe/H] $>$ $-$0.2 in
Figure 7 and $<$ $-$0.2 in Figure 8.  The evolutionary tracks that cover our
range of mass of 0.9 - 1.1 M$_\odot$ are shown for the respective values of
[Fe/H].  Isochrones for 100, 700, 2500 and 4500 myr are also shown in each
figure.  The four sizes of symbols indicates the size of A(Be).

In Figure 8, two stars appear to be $>$1.1 M$_{\odot}$ for this survey, but
this is due to the metallicity dependence of the evolutionary tracks; in LR04
smaller intervals were chosen for the tracks and therefore better constraints
on the mass in LR04 were made.  Although stars which fall near the cutoff
between the two tracks in this study do not appear to be in the right mass
range, the masses for these stars are indeed in the correct range and have
been adopted from the LR04 paper.

As can be seen in Figure 1, many of the stars in our sample have evolved away
from the zero age main sequence (which is the reason why evolutionary ages can
be assigned).  In S04 the cut-off between dwarfs and subgiants is given as log
g = 4.1.  There are 17 stars in this sample that accordingly would be
classified as subgiants.  These stars generally have lower Be abundances with
a mean A(Be) of 0.82 (excluding the two with upper limits).  The stars with
log g $\leq$ 4.1 separate out in the distribution shown in the Be histogram in
the lower portion of Figure 6; the lower three bins have 3, 8, and 4 stars
that are subgiants.  

The evolutionarily older subgiants show typically lower Be abundances at a
given T$_{\rm eff}$ than the dwarf stars.  The older age of the subgiants
allow them to have had more time for slow mixing of the light elements,
leading to depletion on the main sequence while the more advanced evolutionary
stage allows them to have enlarged the surface convection zones leading to
light element dilution.

It may also just be a reflection of small number statistics; S04 do not see a
similar depletion in their subgiants as seen in their Figure 4.  Indeed there
would not be as much Be dilution in evolved stars compared to Li dilution.
Figures 7 and 8 show little sign of a correlation between Be abundance and
evolutionary age along the mass tracks.

\subsection{Trends with Temperature}

This solar-mass sample was selected from a survey of F and G stars in the
local thin disk.  This resulted in a temperature range for our stars from 5600
to 6400 K.  When plotted against temperature, an upper envelope for the Be
abundance becomes apparent as seen in Figure 9.  The highest Be abundances are
present near 5800-600 K, with reduced amounts of Be in the hotter stars of our
sample, 6300-6500 K, in the Li-Be dip region.  For the Hyades this dip region
is about T$_{\rm eff}$ $\sim$ 6300 - 6800 K (e.g. Boesgaard \& Tripicco 1986)
but for older stars the dip region may enlarge toward cooler temperatures
(e.e. Michaud 1986) and show a larger spread in A(Li) (e.g.~Jones et
al.~1999).

The lowest Be {\it detections} for this sample are A(Be) $\sim$0.5 dex and the
Be abundances range up to 1.42.  This order of magnitude spread can be seen
particularly in the cooler stars (i.e. T$_{\rm eff} \leq$ 5900 K).  The stars
in the S04a paper have T$_{\rm eff}$ in the range of 4800 to 6300 K.  They
find a abundance peak in A(Be) near 6100 K, with a gentle decline toward lower
temperatures approaching 5000 K.  Their sample includes planet-host stars,
many of which are metal-rich, whereas our sample has only two stars with
[Fe/H] $>$ 0.00.  Those metal-rich stars show enhanced Be.  Although our
sample shows no such Be abundance peak and decline we note that our sample
covers a specific mass range and thus a smaller range in temperature than that
of S04a; we would not expect to see the decline in Be with decreasing
temperature that they found.

Both slow mixing and convection zone depth vary with the stellar metallicity
at a given mass or T$_{\rm eff}$.  There is evidence of a correlation between
[Fe/H] and A(Be) (e.g.~Boesgaard et al. 2008, Boesgaard \& Novicki 2006, and
$\S$4.4 below).  That correlation needs to be considered when evaluating the
temperature relation.  The range for metallicity for our stars is $-$0.65
$\leq$ [Fe/H] $\leq$ +0.11.  In Figure 10 we show the Be abundances normalized
to Fe as [Be/Fe] versus T$_{\rm eff}$.  (We used the meteoritic A(Be) for the
solar system normalization.)  The normalized ratio, [Be/Fe], shows
some spread at a given temperature, but no dramatic trend with temperature.
And unlike Figure 9, the dwarfs and subgiants show similar relations with
temperature.  This plot of [Be/Fe] against T$_{\rm eff}$ does not yield the
same trend as seen in S04a.

\subsection{Trends with [Fe/H]}

The stars of this sample are members of the thin disk and generally have
considerably higher metallicities than those of the thick disk and halo
populations.  These stars are of metallicity [Fe/H] $>$ $-$0.65, with two
positive metallicity stars.  However, when the Be abundance is plotted against
[Fe/H] in Figure 11, the metal-rich stars of this survey match the trend with
the halo stars sampled in Boesgaard et al.~(2008) remarkably well.  The slope
of the relation between A(Be) and [Fe/H] is 0.86 $\pm$0.02.  The most
metal-rich star (at [Fe/H] = 0.11) in our sample is HD 199960 whose Be
abundance is equivalent to that of the meteoritic value of 1.42 dex (Grevesse
\& Sauval 1998).  This value is also the highest Be abundance in this sample,
supporting the idea that A(Be) increases with [Fe/H].

  This is a remarable
trend, extending over nearly four orders of magnitude in [Fe/H] and over three
orders of magnitude in A(Be).

Another illustration of this relation is to sort the sample into metallicity
bins.  We created four metallicity bins and took the mean A(Be) from the three
most Be-rich stars per bin.  Table 5 shows again that A(Be) increases as
[Fe/H] increases.  LR04 created similar [Fe/H] bins for their data with the Li
abundance and found that as the [Fe/H] increased the Li abundance increased
(which was expected), and so did the mass.  Given that our data were selected
based on mass, it was not possible to see the mass trend; however, we did
apply the same [Fe/H] bins to the 157 LR04 solar mass stars that had Li
abundances.  A(Li) is shown to increase with increasing [Fe/H] as well, but it
is less obvious in our smaller data set.

\subsection{Trends with Age}

One of the aims of this study was to understand the relationship between
A(Be)and age.  The mass constraint was placed at 1 $\pm$ 0.1 M$_{\odot}$ in
order to parameterize the A(Be)-age relation.  Of our stars, all but one are
older than the 4.5 Gyr Sun, eight (or 15\%) have ages $<$ 6 Gyr and two (4\%)
have ages $<$ 5 Gyr.  This fits well with the larger sample of LR04 stars with
masses between 0.9 and 1.1 M$_{\odot}$.  The full LR04 sample includes 451
stars, with 157 (or 35\% of the sample) within our mass constraints; of those
157 stars, only 34 (22\%) have ages $<$ 6 Gyr and only 6 (4\%) have ages $<$5
Gyr.

Figure 12 plots A(Be) against age and [Be/Fe] against age.  The spread in
A(Be) (including upper limits) is nearly 1.5 dex for the entire range of ages;
including only the Be detections, the spread is still 0.9 dex.  It should be
noted that only one star (HD 199960) has larger A(Be) than the Sun with the
rest being below the solar A(Be) value of $\sim$ 1.3 dex (e.g. Boesgaard et
al.~2003).  These stars, which have evolved away from the zero age main
sequence, have continued to lose Be.  In the lower panel one can see that the
scatter is reduced by using [Be/Fe], correcting for each star's metallicity.
A small trend is found such that the older stars have more Be relative to Fe
in our one-solar mass sample.  The ratio of Be/Fe is two times larger in the
stars of 10 Gyr compared to those of 5 Gyr.

\subsection{Trends with Lithium Abundance}

Understanding the correlation between Be and Li is key to understanding the
nature of light element depletion and slow mixing in these stars.  Figure 13
shows the plot of A(Be) versus A(Li).

B05 found a linear relationship between A(Li) and A(Be) for field and cluster
stars in the T$_{\rm eff}$ ranges of 5900-6300 K (the Li ``plateau'') and
6300-6650 K (the cool side of the Li-dip) with slopes of 0.365$\pm$0.036 and
0.404$\pm$0.034 respectively.  We can combine our data with the field stars of
B04 in the temperature range of 5900 $<$ T$_{\rm eff}$ $<$ 6300 K.  Figure 13
shows those results for the total of 60 stars.  The triangles are stars from
this study.  For the combined data we find a slope of 0.34 $\pm$0.05 which
compares well with the slope for 31 field stars in this same temperature range
in B04: 0.36 $\pm$0.05.  Note that the three most Li- and Be-depleted stars in
Figure 13 all have solar metallicity: [Fe/H] = -0.03, +0.09, +0.12.  It
appears that metallicity is not an important factor in the Li-Be correlation.

\subsection{The Spread in Be Abundance}

There is clear evidence that astration of Be is taking place in these
one-solar-mass stars.  The spread in A(Be) and [Be/Fe] at a given temperature
shows that.  The spread in A(Be) covers more than 1.5 dex, while the typical
errors are $\pm$0.15.  There are two stars in our sample that have nearly
identical stellar parameters, yet they differ in both A(Be) and A(Li) by a
factor of two.  The Be region of the spectra of these two stars, HD 200580 and
HD 204712, are shown in Figure 14.  Only the two Be II lines show meaningful
differences between these two spectra.  Their full set of stellar parameters
are given in Table 4.  They differ in T$_{\rm eff}$ by only 35 K, in log g by
0.07, and in [Fe/H] by only 0.06, all within the precision of their
determinations.  Their respective masses are 0.95 and 0.96 M$_{\odot}$ and
their ages are 9.17 and 9.35 Gyr, with the slightly older one having more Be.

This pair may indicate that another parameter plays a role in Be astration; it
has been suggested by B04a that rotational velocity is that
parameter.  Stars are formed with differing initial velocities and spin down
to their present-day values.  During the spin-down, extra mixing occurs which
increases the amount of Be (and Li) that can reach the critical temperature
for destruction.  As found by B04a the depletion of Li and Be observed in
cluster and field stars is well-matched by the rotation-model predictions of
Deliyannis \& Pinsonneault (1997).  See Figure 12 of B04a.

\section{SUMMARY AND CONCLUSIONS}

We have determined Be abundances for 50 solar-mass stars using spectra taken
at the Keck I telescope and the CFH Telescope.  The stars were selected from
LR04 to be within 0.1 M$_{\odot}$ of the sun.  Most of our sample of
one-solar-mass stars were taken using the new Keck I HIRES mosaic CCD, the
resolution was $\sim$78,000 with a median S/N ratio of 146.  The data were
reduced using the IDL HIRES Redux pipeline and IRAF.  We adopted the stellar
parameters as determined by LR04 and used them to create model atmospheres for
spectral synthesis in MOOG using the Kurucz grid point models.  Be abundances
were determined by fitting the 4 \AA{} region around the Be II resonance lines
located at 3130.421 and 3131.065 \AA{}.  We plotted HR diagrams using the G00
isochrones and evolutionary tracks in combination with the LR04 stellar
parameters.  The majority of our stars are evolved away from the zero age main
sequence and all but one are older than the Sun.

The main results from our analysis are as follows:

In spite of the similarity in mass of our stars, there is a wide range in Be
abundances from A(Be) $<$ $-$0.2 to +1.42, or more than a factor of 40.  The
median Be abundance is 0.92 -- for the 46 stars with Be detections -- which
represents a deficiency relative to the meteoritic abundance of A(Be) = 1.42.
The median Li abundance for the 44 stars with Li detections is A(Li) = 2.39,
also deficient compared to the meteoritic value of 3.31.  See Figure 6 for
the distribution of light elements in our stars.  Of our four stars with upper
limits on the Be abundances, all have upper limits on Li also.  This is
expected since Li is destroyed in stellar interiors at lower temperatures than
Be by (p,$\alpha$), making Li more susceptible to destruction.

We have compared the positions of our stars on the HR diagram with the
isochrones and evolutionary tracks of G00 (see Figures 7 and 8).  In our
sample of 50 stars the subgiants, as defined by log g $<$4.1, make up
one-third of the stars.  At a given T$_{\rm eff}$ these more-evolved stars
generally have lower Be abundances than the main-sequence stars.  The reduced
amount of Be could result both from more depletion on the main sequence and
possibly from more dilution due to the expansion of the surface convection
zone in the evolutionarily more advanced subgiants.

This one-solar-mass sample of thin disk stars matches the remarkable
relationship between A(Be) and [Fe/H] found earlier (e.g. Boesgaard et
al.~1999, Boesgaard \& Novicki 2006, Boesgaard et al.~2008).  As can be seen
in Figure 11, this relationship extends over 4 orders of magnitude in [Fe/H]
and more than 3 orders of magnitude in A(Be).  The linear relation between
these two logarithmic quantities has a slope of 0.86 $\pm$0.02.  This
indicates an increase in the abundance of both elements over the age of the
Galaxy even though Fe is made by nuclear processing in stars and Be is made
through spallation reactions outside of stars.

In the relation between A(Be) and T$_{\rm eff}$ there are lower Be abundances
near the F star Li-Be dip, while the higher Be abundances appear in the cooler
stars, as can be seen in Figure 9.  In fact, there is a large spread in A(Be)
in the cooler stars.  Due to the relationship between A(Be) and [Fe/H], we
have normalized the Be abundances to the Fe abundances and form [Be/Fe].  When
this parameter is plotted against T$_{\rm eff}$ (see Figure 10), there is no
temperature dependence.  The main sequence stars and the subgiants have
similar distributions of [Be/Fe] with T$_{\rm eff}$.  Even with this
normalization, however, there is a large spread in [Be/Fe] of a factor of 6
(not including the upper limits on Be).

The spread that we find is larger than the typical error of $\sim$0.12 dex in
the Be abundances.  We can see this in the pair of nearly identical stars, HD
200580 and HD 204712, which differ in Be abundance by a factor of 2
although they have the same [Fe/H], T$_{\rm eff}$, log g, age and mass.  See
Figure 14.

The range in age of our stars in the one-solar-mass sample covers 3.8 to 9.9
Gyr.  In this age range there is a spread of 1.5 dex in A(Be).  However, as
can be seen in Figure 12, there is no noticeable trend if Be abundance with
age.  All except one of our stars are older than the sun and all, except one,
have lower Be abundances than that found in meteorites (and in the Sun).  The
stellar deficiencies in Be are probably indicative of the longer time over
which slow mixing and depletion of Be has taken place.  For the stars that
have evolved toward the base of the red giant branch, the deepening of the
surface convection zone could have produced additional dilution of the surface
Be content.

Of our 50 stars, 30 are in the temperature regime of T$_{\rm eff}$ = 5900 -
6300 K.  This range corresponds to the ``Li plateau'' region in the Hyades
cluster (see B04a).  This doubles the number of field stars with detectable Li
and Be and confirms the linear correlation of A(Li) and A(Be).  The slope is
+0.34 $\pm$0.05 for the 60 stars in this temperature region for main-sequence
and subgiant stars.  The Li depletion is more severe than the Be depletion, as
expected.  The observed characteristics are in excellent agreement with the
predictions of rotational mixing by Deliyannis \& Pinsonneault (1997).

\acknowledgements 
We are grateful to the Keck Observatory support astronomers,
Grant Hill, Jeffrey Mader, Hien Tran, and Greg Wirth for their assistance with
HIRES and to Megan Novicki and Jeffrey Rich for help in the data reduction.
JAK was supported through NSF-Research Experiences for Undergraduates program
to the University of Hawaii grant AST 04-53395.  This work was supported by
NSF AST 05-05899 to AMB.

\clearpage

\singlespace
\begin{center}
\begin{deluxetable}{lccrcc} 
\tablewidth{0pc}
\tablecolumns{6} 
\tablecaption{Keck HIRES Observations} 
\tablehead{ 
\colhead{Star HD} &  \colhead{Star HR} &  \colhead{V}  &  \colhead{Night} & \colhead{Exp. Time} & \colhead{Total}  \\
\colhead{} & \colhead{} & \colhead{} & \colhead{} & \colhead{(min)} & \colhead{S/N}
} 
\startdata 

HD 5494  & \nodata &  7.95 & 02 Jan 2006 & \phn7 & 267\\
HD 14877 & \nodata &  8.46 & 31 Jan 2005 & \phn8 & 107\\
HD 22521 & \nodata &  6.95 & 31 Jan 2005 & \phn4 & 146\\
HD 24421 & \nodata &  6.83 & 31 Jan 2005 & \phn4 & 148\\
HD 25173 & \nodata &  7.16 & 31 Jan 2005 & \phn8 & \phn53\\
HD 26421 & \nodata &  8.19 & 31 Jan 2005 & 10 & 127\\
HD 28620 & \nodata &  6.81 & 31 Jan 2005 & \phn8 & 226\\
HD 30743 & HR 1545 &  6.26 & 27 Sep 2005 & \phn3 & 208\\
HD 33632 & \nodata &  6.46 & 31 Jan 2005 & \phn6 & 226\\
HD 35296 & HR 1780 &  5.00 & 11 Jan 2007 & 20 & \phn40\\
HD 54717 & \nodata &  7.20 & 01 Apr 2005 & 25 & \phn71\\
HD 63333 & \nodata &  7.11 & 01 Apr 2005 & 20 & \phn47\\
HD 68284 & \nodata &  7.78 & 01 Apr 2005 & 60 & \phn76\\
HD 78418 & HR 3626 &  5.98 & 11 Jan 2007 & \phn5 & 115\\
HD 80218 & \nodata &  6.63 & 01 Apr 2005 & 15 & \phn78\\
HD 87838 & \nodata &  7.73 & 11 Jan 2007 & 30 & \phn59\\
HD 88446 & \nodata &  7.85 & 11 Jan 2007 & 30 & \phn33\\
HD 89125 & HR 4039 &  5.80 & 01 Apr 2005 & 15 & 204\\
HD 91638 & \nodata &  6.68 & 01 Apr 2005 & 25 & 133\\
HD 91889 & HR 4158 &  5.70 & 02 Jan 2006 & \phn4 & 350\\
HD 94012 & \nodata &  7.85 & 11 Jan 2007 & 30 & 113\\
HD 94835 & \nodata &  7.96 & 01 Apr 2005 & 30 & 252\\
HD 100180 & HR 4437 & 6.20 & 01 Apr 2005 & 20 & 330\\
HD 101676 & \nodata & 7.09 & 11 Jan 2007 & 25 & 197\\
HD 108510 & \nodata & 6.69 & 11 Jan 2007 & 25 & 260\\
HD 108845 & HR 4761 & 6.21 & 09 Jun 2007 & \phn3 & 273\\
HD 109303 & \nodata & 8.15 & 09 Jun 2007 & 15  & 289\\
HD 112756 & \nodata & 8.15 & 11 Jan 2007 & 20 & 126\\
HD 118244 & \nodata & 6.99 & 02 Jan 2006 & \phn5 & 235\\
HD 121560 & HR 5243 & 6.10 & 11 Jan 2007 & \phn7 & 146\\
HD 169359 & \nodata & 7.81 & 06 Jul 2005 & \phn7 & 110\\
HD 186379 & \nodata & 6.76 & 27 Sep 2005 & \phn5 & 257\\
HD 200580 & \nodata & 7.33 & 09 Jun 2007 & \phn6 & 277\\
HD 202884 & \nodata & 7.27 & 18 Nov 2004 & \phn4 & 122\\
HD 204712 & \nodata & 7.65 & 09 Jun 2007 & \phn6 & 252\\
HD 209320 & \nodata & 8.32 & 02 Jan 2006 & \phn5 & 150\\
HD 209858 & \nodata & 7.79 & 18 Nov 2004 & 10 & 148\\
HD 215442 & \nodata & 7.53 & 18 Nov 2004 & \phn7 & 124\\
HD 217877 & HR 8772 & 6.68 & 05 Jul 2005 & \phn5 & 173\\
HD 221356 & HR 8931 & 6.49 & 05 Jul 2005 & \phn5 & 188\\

\enddata 

\end{deluxetable} 
\end{center}


\clearpage

\singlespace
\begin{center}
\begin{deluxetable}{llcrccll} 
\tablewidth{0pc}
\tablecolumns{8} 
\tablecaption{Observations} 
\tablehead{ 
\colhead{Star} &  \colhead{} &  \colhead{V}  &  \colhead{Night} & \colhead{Exp. Time} & \colhead{Total} & \colhead{Telescope} & \colhead{Ref.\tablenotemark{1}} \\

\colhead{HD} & \colhead{HR} & \colhead{} & \colhead{} & \colhead{(min)} & \colhead{S/N} &\colhead{} & \colhead{} } 

\startdata 
HD 693 & HR 33 & 4.89 & 05 Jan 2002 & \phn5 & 145 & Keck I HIRES & B04a \\
HD 4813 & HR 235 & 5.19 & 16 Oct 1995 & 15 & \phn72 & CFHT Gecko & B01 \\
HD 11592 & \nodata & 6.78 & 16 Oct 1995 & 45 & \phn48 & CFHT Gecko & B04b \\
HD 15335 & HR 720 & 5.88 & 15 Nov 1999 & 10 & 108 & Keck I HIRES & B04a \\
HD 58551 & HR 2835 & 6.54 & 05 Jan 2002 & 12 & 145  & Keck I HIRES & B04a \\
HD 58855 & HR 2849 & 5.37 & 15 Oct 1995 & 20 & \phn88 & CFHT Gecko & B01 \\
HD 103799 & HR 4572 & 6.62 & 04 Jan 2002 & 10 & 140  & Keck I HIRES & B04a \\
HD 142860 & HR 5933 & 3.85 & 16 Jul 1992 &\phn5 & \phn52 & CFHT f/8 Coud\'e & B04b \\
HD 182101 & HR 7354 & 6.35 & 15 Oct 1995 & 30 & \phn73 & CFHT Gecko & B01 \\
HD 199960 & HR 8041 & 6.20 & 13 Nov 1999 & 12 & \phn93 & Keck I HIRES & B04a \\
HD 218470 & HR 8805 & 5.70 & 15 Oct 1995 & 25 & \phn79 & CFHT Gecko & B01 \\
\tablenotetext{1}{B01=Boesgaard et al.~2001, B04a=Boesgaard et al.~2004a,
B04b=Boesgaard et al.~2004b}
\enddata

\end{deluxetable}
\end{center}


\clearpage

\singlespace
\begin{center}
\begin{deluxetable}{lrrcr} 
\tablewidth{0pc}
\tablecolumns{5}
\tablecaption{Stellar Parameters} 
\tablehead{ 
\colhead{Star}  &  \colhead{T$_{\rm eff}$}  &  \colhead{log g} & \colhead{[Fe/H]} & \colhead{$\xi$}}

\startdata 
HD 693 & 6132 & 4.12 & $-$0.48 & 1.85\\
HD 4813 & 6286 & 4.34  & $-$0.15 & 1.58\\
HD 5494 & 6082 & 4.00  & $-$0.17 & 1.97\\
HD 11592 & 6234 & 4.20 & $-$0.35 & 1.83\\
HD 14877 & 5970 & 4.03 & $-$0.42 & 1.84\\
HD 15335 & 5785 & 3.92 & $-$0.22 & 1.83\\
HD 22521 & 5783 & 3.96 & $-$0.25 & 1.78\\
HD 24421 & 5987 & 4.14 & $-$0.38 & 1.71\\
HD 25173 & 5867 & 4.07 & $-$0.62 & 1.70\\
HD 26421 & 5737 & 3.98 & $-$0.39 & 1.72\\
HD 28620 & 6101 & 4.08 & $-$0.52 & 1.88\\
HD 30743 & 6222 & 4.15 & $-$0.62 & 1.88\\
HD 33632 & 5962 & 4.30 & $-$0.23 & 1.48\\
HD 35296 & 6015 & 4.24 & $-$0.14 & 1.60\\
HD 54717 & 6350 & 4.26 & $-$0.44 & 1.84\\
HD 58551 & 6149 & 4.22 & $-$0.54 & 1.73\\
HD 58855 & 6286 & 4.31 & $-$0.31 & 1.73\\
HD 63333 & 6057 & 4.23 & $-$0.39 & 1.65\\
HD 68284 & 5832 & 3.91 & $-$0.56 & 1.88\\
HD 78418 & 5625 & 3.98 & $-$0.26 & 1.63\\
HD 80218 & 6092 & 4.16 & $-$0.28 & 1.77\\
HD 87838 & 6019 & 4.29 & $-$0.43 & 1.54\\
HD 88446 & 5875 & 4.08 & $-$0.52 & 1.70\\
HD 89125 & 6038 & 4.25 & $-$0.36 & 1.61\\
HD 91638 & 6159 & 4.29 & $-$0.25 & 1.65\\
HD 91889 & 6051 & 4.20 & $-$0.28 & 1.68\\
HD 94012 & 6064 & 4.41 & $-$0.47 & 1.42\\
HD 94835 & 5814 & 4.43 & +0.05 & 1.19\\
HD 100180 & 5866 & 4.12 & $-$0.11 & 1.64\\
HD 101676 & 6102 & 4.09 & $-$0.47 & 1.87\\
HD 103799 & 6169 & 4.02 & $-$0.45 & 2.01\\
HD 108510 & 5929 & 4.31 & $-$0.06 & 1.44\\
HD 108845 & 6060 & 4.08 & $-$0.24 & 1.84\\
HD 109303 & 5904 & 4.05 & $-$0.52 & 1.76\\
HD 112756 & 5993 & 4.40 & $-$0.35 & 1.37\\
HD 118244 & 6234 & 4.13 & $-$0.55 & 1.92\\
HD 121560 & 6059 & 4.35 & $-$0.38 & 1.49\\
HD 142860 & 6227 & 4.18 & $-$0.22 & 1.85\\
HD 169359 & 5810 & 4.12 & $-$0.31 & 1.59\\
HD 182101 & 6344 & 4.22 & $-$0.29 & 1.89\\
HD 186379 & 5811 & 3.99 & $-$0.43 & 1.76\\
HD 199960 & 5750 & 4.17 & $+$0.11 & 1.48\\
HD 200580 & 5853 & 4.05 & $-$0.54 & 1.72\\
HD 202884 & 6141 & 4.36 & $-$0.24 & 1.55\\
HD 204712 & 5888 & 4.12 & $-$0.48 & 1.65\\
HD 209320 & 5994 & 4.14 & $-$0.18 & 1.71\\
HD 209858 & 5911 & 4.26 & $-$0.27 & 1.49\\
HD 215442 & 5872 & 3.38 & $-$0.22 & 2.60\\
HD 217877 & 5872 & 4.28 & $-$0.18 & 1.43\\
HD 218470 & 6495 & 4.06 & $-$0.13 & 2.22\\
HD 221356 & 6005 & 4.42 & $-$0.24 & 1.36\\
\enddata 

\end{deluxetable} 
\end{center}


\clearpage

\singlespace
\begin{center}
\begin{deluxetable}{lrrccrrr} 
\tablewidth{0pc}
\tablecolumns{7}
\tablecaption{Stellar Parameters and Abundances} 
\tablehead{ 
\colhead{Star}  &  \colhead{[Fe/H]}  &  \colhead{T$_{\rm eff}$} & \colhead{M$_V$}& \colhead{Mass} & \colhead{Age} & \colhead{A(Li)} & \colhead{A(Be)}}

\startdata

HD 693 & $-$0.48  & 6132 & 3.51 & 1.04  & 6.11  & 2.39  & 0.88\\ 
HD 4813   & $-$0.15  & 6146 & 4.23 & 1.07  & 3.81  & 2.74  & 1.07\\
HD 5494   & $-$0.17  & 6082 & 3.22& 1.04  & 8.85  &$<$1.04  & 0.57\\
HD 11592  & $-$0.35  & 6234 & 3.66 & 1.05  & 5.77  & 2.33  & 0.87\\
HD 14877  & $-$0.42  & 5970 & 3.67 & 0.98  & 8.34  & 2.39  & 0.77\\ 
HD 15335  & $-$0.22  & 5785 & 3.45 & 1.08  & 6.82  & 2.39  & 1.07\\
HD 22521  & $-$0.25  & 5783 & 3.60 & 1.03  & 8.10  & 2.53  & 1.00\\
HD 24421  & $-$0.38  & 5987 & 3.81 & 0.98  & 8.46  & 2.56  & 1.00\\
HD 25173  & $-$0.62  & 5867 & 3.52 & 0.96  & 8.63  & 2.49  & \nodata \\
HD 26421  & $-$0.39  & 5737 & 3.68 & 0.98  & 9.02  & 1.91  & 0.87\\
HD 28620  & $-$0.52  & 6101 & 3.61 & 0.98  & 7.75  & 2.40  & 0.70\\
HD 30743  & $-$0.62  & 6222 & 3.53 & 1.00  & 6.54  & 2.35  & 0.81\\
HD 33632  & $-$0.23  & 5962 & 4.40 & 0.97  & 7.59  & 2.45  & 0.90\\
HD 35296  & $-$0.14  & 6015 & 4.15 & 1.02  & 7.12  & 2.87  & 0.97\\ 
HD 54717  & $-$0.44  & 6350 & 3.86 & 1.04  & 5.34  & 2.34  & 0.69\\
HD 58551  & $-$0.54  & 6149 & 4.09 & 0.92  & 9.14  & 2.49  & 0.81\\ 
HD 58855  & $-$0.31  & 6286 & 3.87 & 1.07  & 5.00 & 2.22  & 0.92\\
HD 63333  & $-$0.39  & 6057 & 3.92 & 0.97  & 8.52  & 2.38  & 0.85\\
HD 68284  & $-$0.56  & 5832 & 3.41 & 1.00  & 7.22  & 2.39  & 0.74\\
HD 78418  & $-$0.26  & 5625 & 3.48 & 1.07  & 6.73  & 2.03  & 1.00\\
HD 80218  & $-$0.28  & 6092 & 3.65 & 1.04  & 6.87  & $<$1.02  & $<$-0.18\\
HD 88446  & $-$0.52  & 5875 & 3.68 & 0.94  & 9.62  & 0.85  & 0.52\\
HD 87838  & $-$0.43  & 6019 & 4.21 & 0.91  & 10.2  & 2.37  & 1.12\\
HD 89125  & $-$0.36  & 6038 & 4.03 & 0.98  & 7.97  & 2.38  & 0.99\\
HD 91638  & $-$0.25  & 6159 & 3.93 & 1.02  & 7.07  & 2.55  & 1.02\\
HD 91889  & $-$0.28  & 6051 & 3.76 & 1.03  & 7.05  & 2.47  & 0.88\\
HD 94012  & $-$0.47  & 6064 & 4.38 & 0.90  & 9.78  & 2.46  & 1.02\\
HD 94835  &  +0.05   & 5814 & 4.47 & 1.02  & 6.88  & 2.11  & 1.24\\
HD 100180 & $-$0.11  & 5866 & 4.44 & 0.98  & 8.01  & 2.43  & 0.92\\
HD 101676 & $-$0.47  & 6102 & 3.64 & 0.99  & 7.83  & 2.38  & 0.84\\
HD 103799 & $-$0.45  & 6169 & 3.29 & 1.07  & 6.06  & 2.18  & 0.87\\
HD 108510 & $-$0.06  & 5929 & 4.44 & 1.02  & 6.45  & 2.39  & 1.27\\
HD 108845 & $-$0.24  & 6060 & 2.94 & 1.08  & 7.25  & 1.98  & 0.83\\
HD 109303 & $-$0.52  & 5904 & 3.55 & 0.98  & 8.46  & $<$1.65 & 0.85\\
HD 112756 & $-$0.35  & 5993 & 4.52 & 0.92  & 8.99  & 2.46  & 1.10\\
HD 118244 & $-$0.55  & 6234 & 3.71 & 0.97  & 7.77  & 2.07  & 0.74\\
HD 121560 & $-$0.38  & 6059 & 4.24 & 0.95  & 8.70  & 2.43  & 1.01\\
HD 142860 & $-$0.22  & 6227 & 3.63 & 1.09  & 5.53  & 2.15  & 0.82\\
HD 169359 & $-$0.31  & 5810 & 3.91 & 0.96  & 9.71  & 2.33  & 0.87\\
HD 182101 & $-$0.29  & 6344 & 3.58 & 1.09  & 5.11  & $<$1.88  & $<$0.65\\
HD 186379 & $-$0.43  & 5811 & 3.60 & 0.99  & 8.06  & 2.30  & 1.02\\
HD 199960 & +0.11    & 5750 & 4.10 & 1.08  & 6.48  & 2.37  & 1.42\\
HD 200580 & $-$0.54  & 5853 & 3.57 & 0.96  & 9.17  & 2.08  & 0.65\\
HD 202884 & $-$0.24  & 6141 & 4.15 & 1.03  & 5.57  & 2.63  & 0.92\\
HD 204712 & $-$0.48  & 5888 & 3.71 & 0.95  & 9.35  & 2.42  & 0.95\\
HD 209320 & $-$0.18  & 5994 & 3.78 & 1.05  & 6.87  & 2.45  & 1.02\\ 
HD 209858 & $-$0.27  & 5911 & 4.06 & 0.97  & 8.88  & 2.36  & 0.96\\
HD 215442 & $-$0.22  & 5872 & 2.96 & 1.00  & 9.86  &$<$0.85  & $<$0.30\\
HD 217877 & $-$0.18  & 5872 & 4.24 & 0.97  & 9.24  & 2.16  & 1.14\\
HD 218470 & $-$0.13  & 6495 & 3.04 & 1.09  & 7.66  &$<$1.52  & $<$0.1\\
HD 221356 & $-$0.24  & 6005 & 4.41 & 0.98  & 6.44  & 2.50  & 1.10\\

\enddata

\end{deluxetable}
\end{center}

\clearpage

\singlespace
\begin{center}
\begin{deluxetable}{lcc} 
\tablewidth{0pc}
\tablecolumns{3}
\tablecaption{Mean Be abundances calculated from the three most Be-rich and
mean Li abundances from the six most Li-rich stars in each [Fe/H] bin.} 
\tablehead{
\colhead{[Fe/H]}  &  \colhead{A(Be)} &  \colhead{A(Li)}
}
\startdata
+0.2 to $-$0.1 & 1.31   & 2.58 \\
$-$0.1 to $-$0.3 & 1.10 & 2.91 \\
$-$0.3 to $-$0.5 & 1.08 & 2.63 \\
$-$0.5 to $-$0.7 & 0.82 & 2.49 \\

\enddata

\end{deluxetable} 
\end{center}


\clearpage

\begin{figure}
\plotone{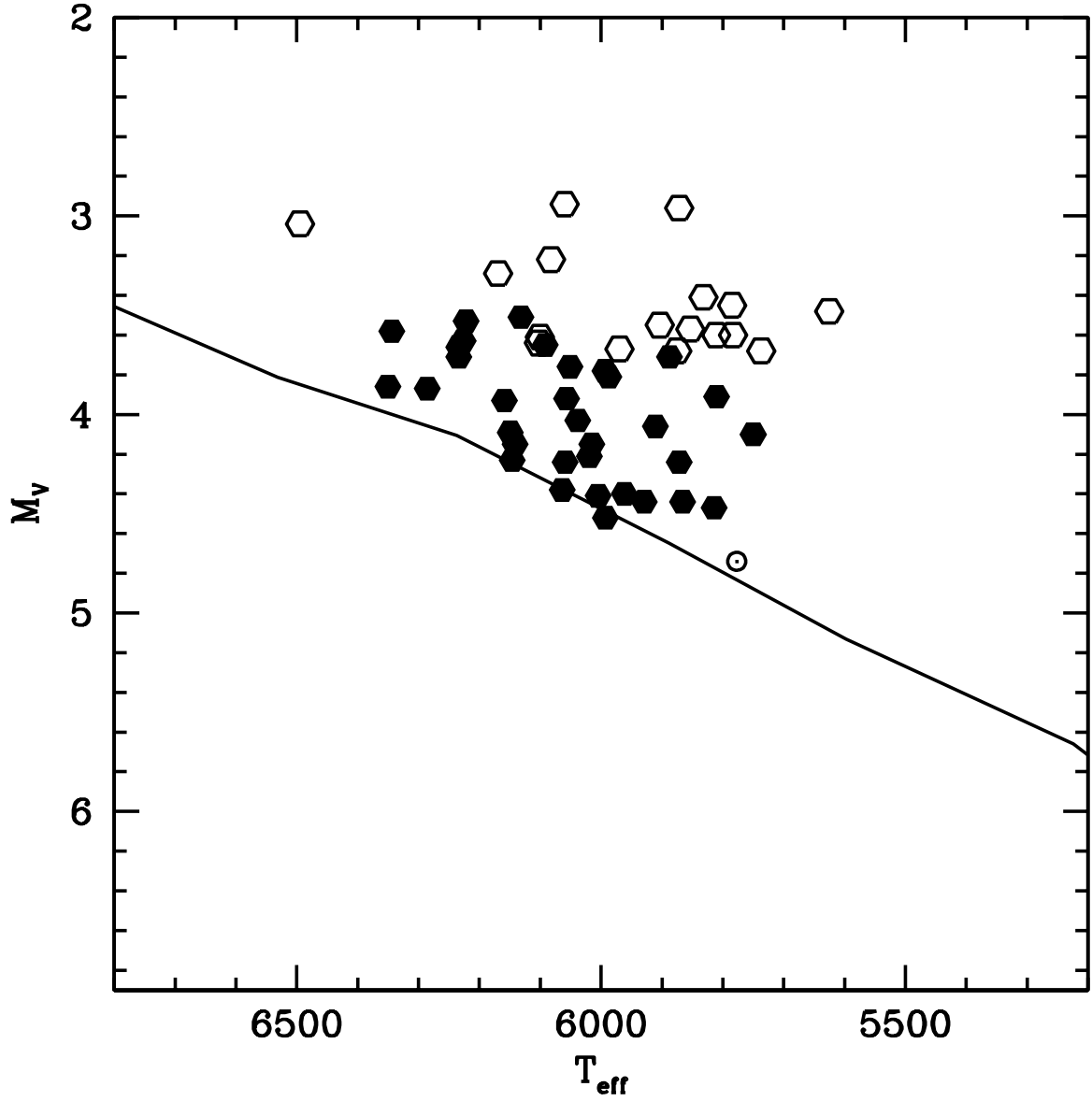}
\caption{HR Diagram of the program stars plotted with the zero-age main
sequence (from Girardi et al.~2000).  The dwarf stars (log g $>$4.1) are filled
hexagons and the subgiants are open hexagons.  The dotted circle represents
the Sun.}
\end{figure}

\begin{figure}
\plotone{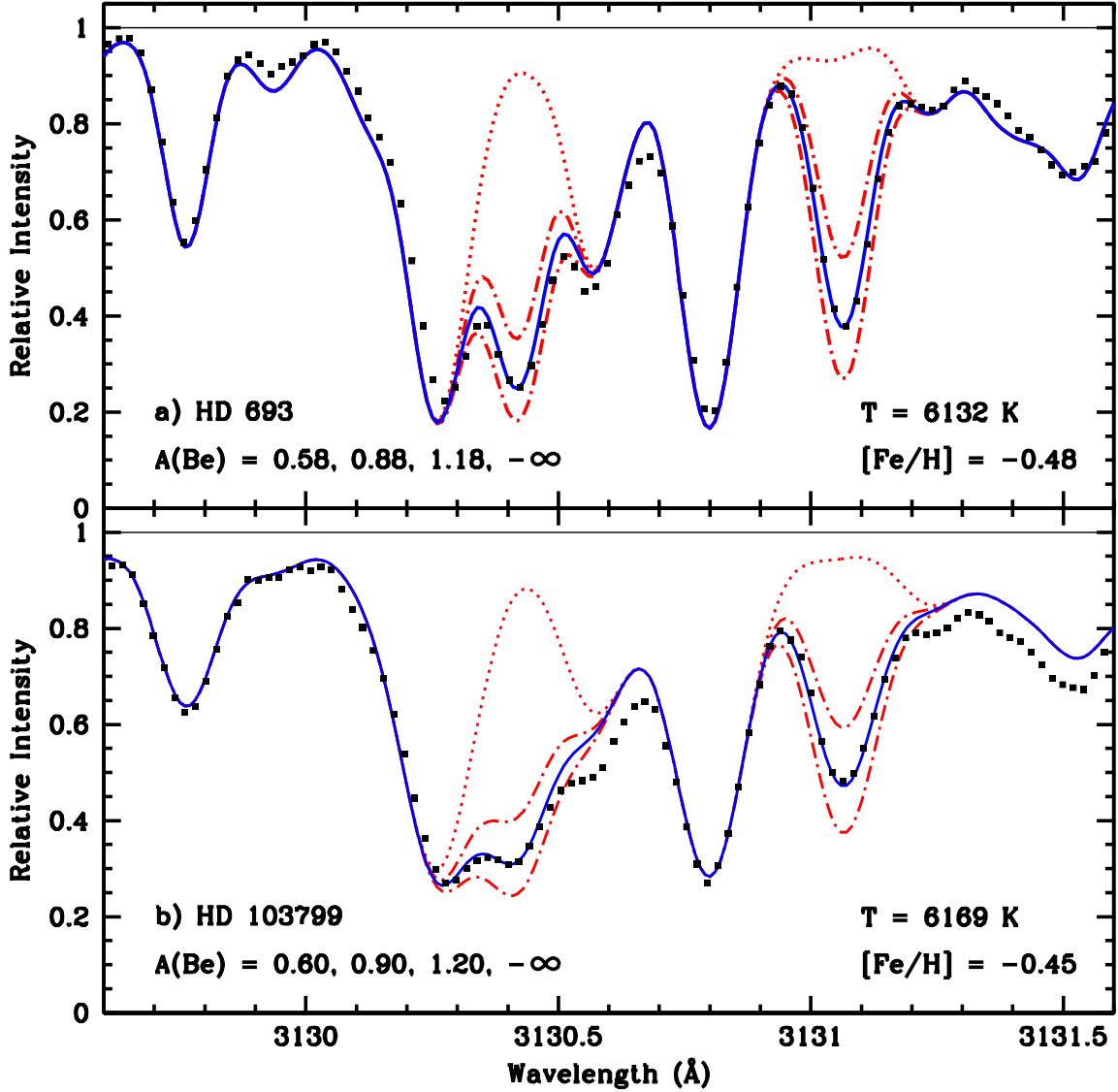}
\caption{Figures 2-5 show spectral syntheses of the Be II region for selected
stars of our sample.  The data (points) and best fit (solid line) are shown,
as well as the fits for a factor of two above and below the best fit for A(Be)
(dashed-dotted lines) and the fit for a star with zero Be.  HD 693 and HD
103799 have the same A(Be), T$_{\rm eff}$, [Fe/H], and age, but differ in
rotation with HD 693 having a FWHM gaussian at 0.09 and HD 103799 at 0.13}
\end{figure}

\begin{figure}
\plotone{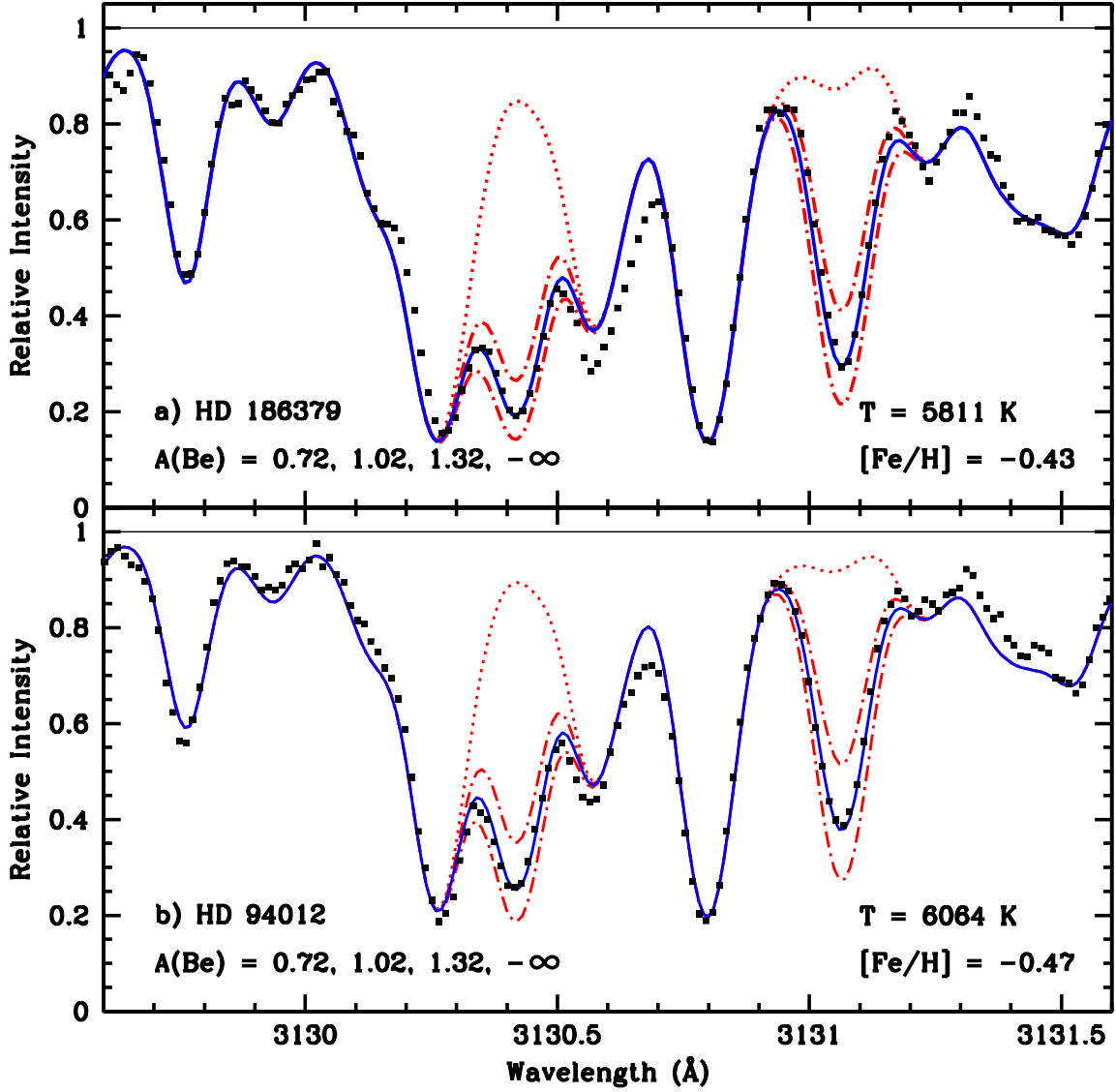}
\caption{These stars are of similar A(Be) and [Fe/H], but differ in age (10
Gyr vs 8 Gyr) and T$_{\rm eff}$ by 250 K.  The Be abundance is the same in
spite of the differences in age and T$_{\rm eff}$.}
\end{figure}

\begin{figure}
\plotone{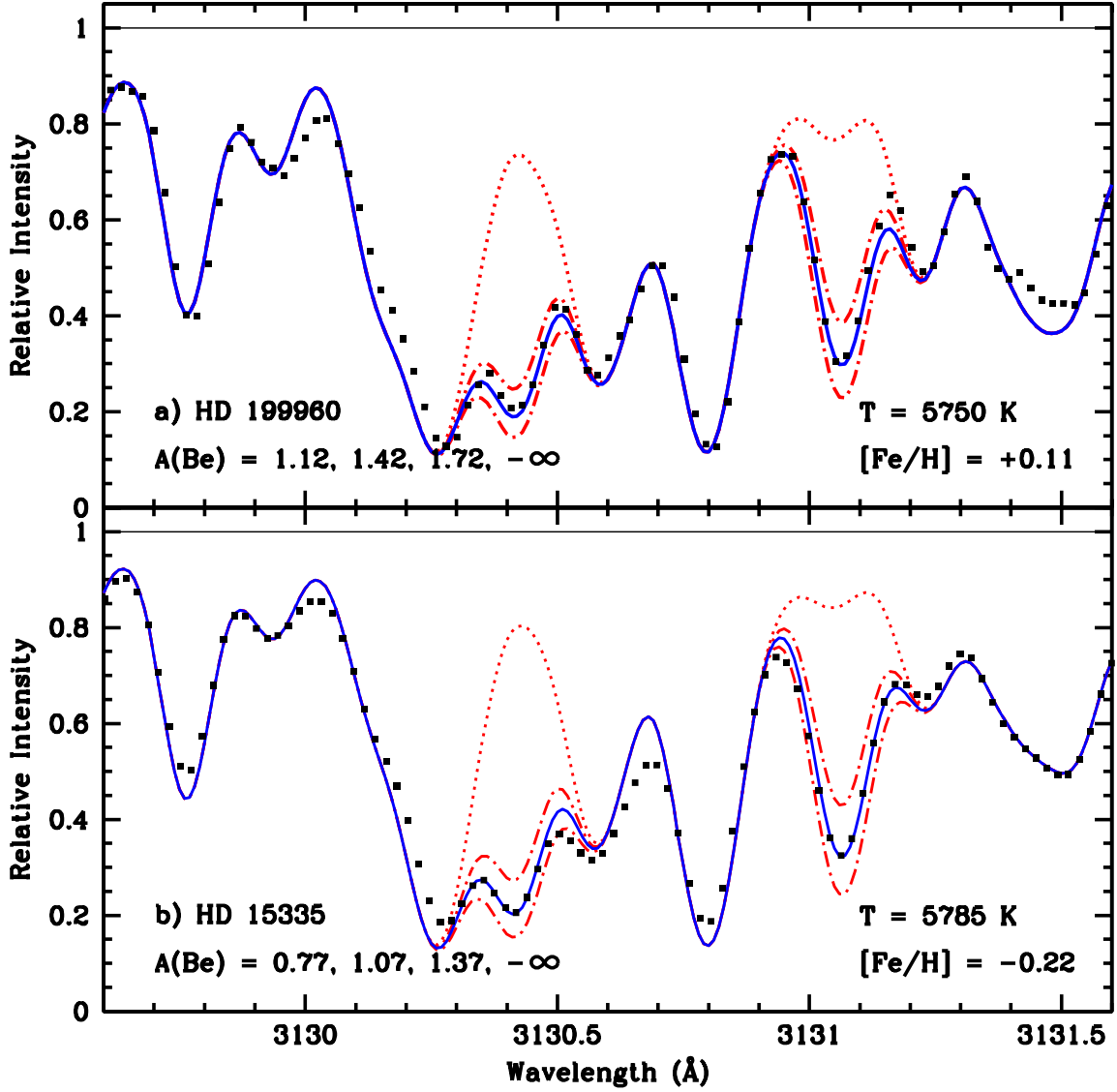}
\caption{These solar temperature stars are the same age.  Both A(Be) and
[Fe/H] are higher by a factor of $\sim$2 in HD 199960 which is the most Be-rich
star in this sample.}
\end{figure}

\begin{figure}
\plotone{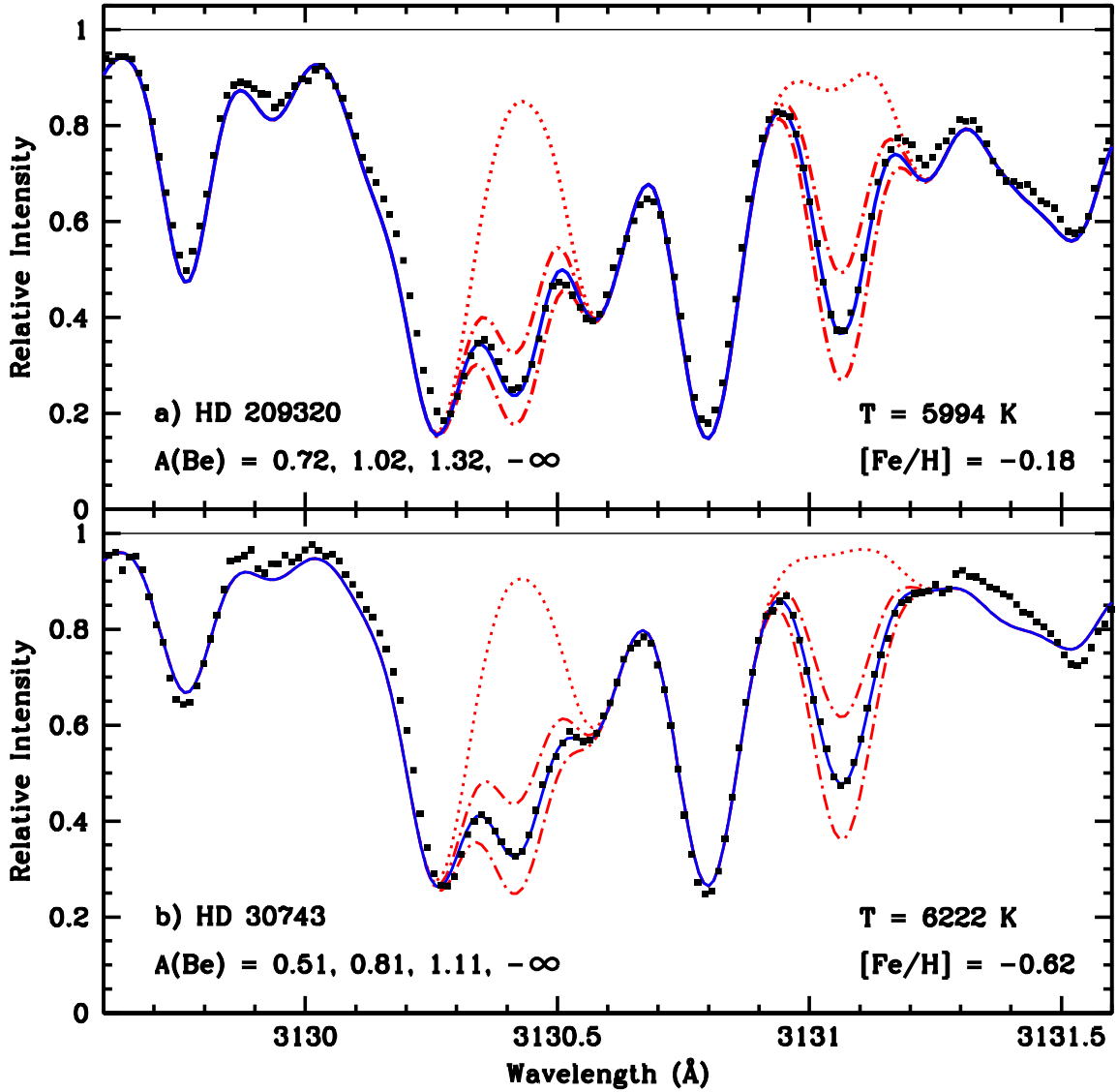}
\caption{These stars have the same age at $\sim$6.6 Gyr, but differ in T$_{\rm
eff}$, [Fe/H], and A(Be).  Iron and Be are higher by a factor of 2.7 and 1.6
respectively in HD 209320.  This corresponds well to the trend shown in Figure
11.}
\end{figure}

\begin{figure}
\plotone{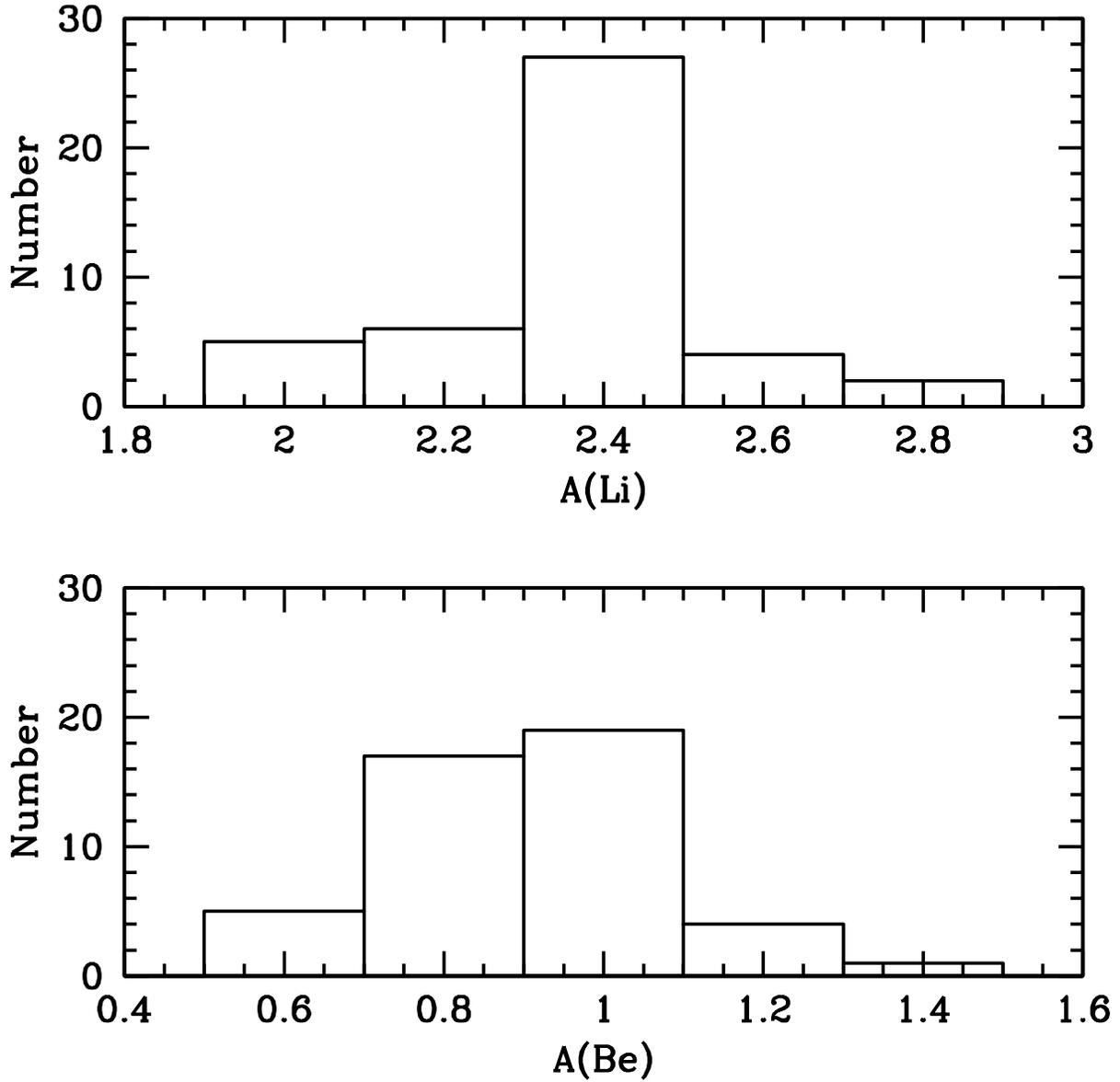}
\caption{Histograms of the abundances of Li (top) and Be (bottom).  For the 44
stars with Li detections the median A(Li) is 2.39 while for the 46 stars with
Be detections the median is A(Be) is 0.92.  The meteoritic values are 3.31 and
1.42 respectively.}
\end{figure}

\begin{figure}
\plotone{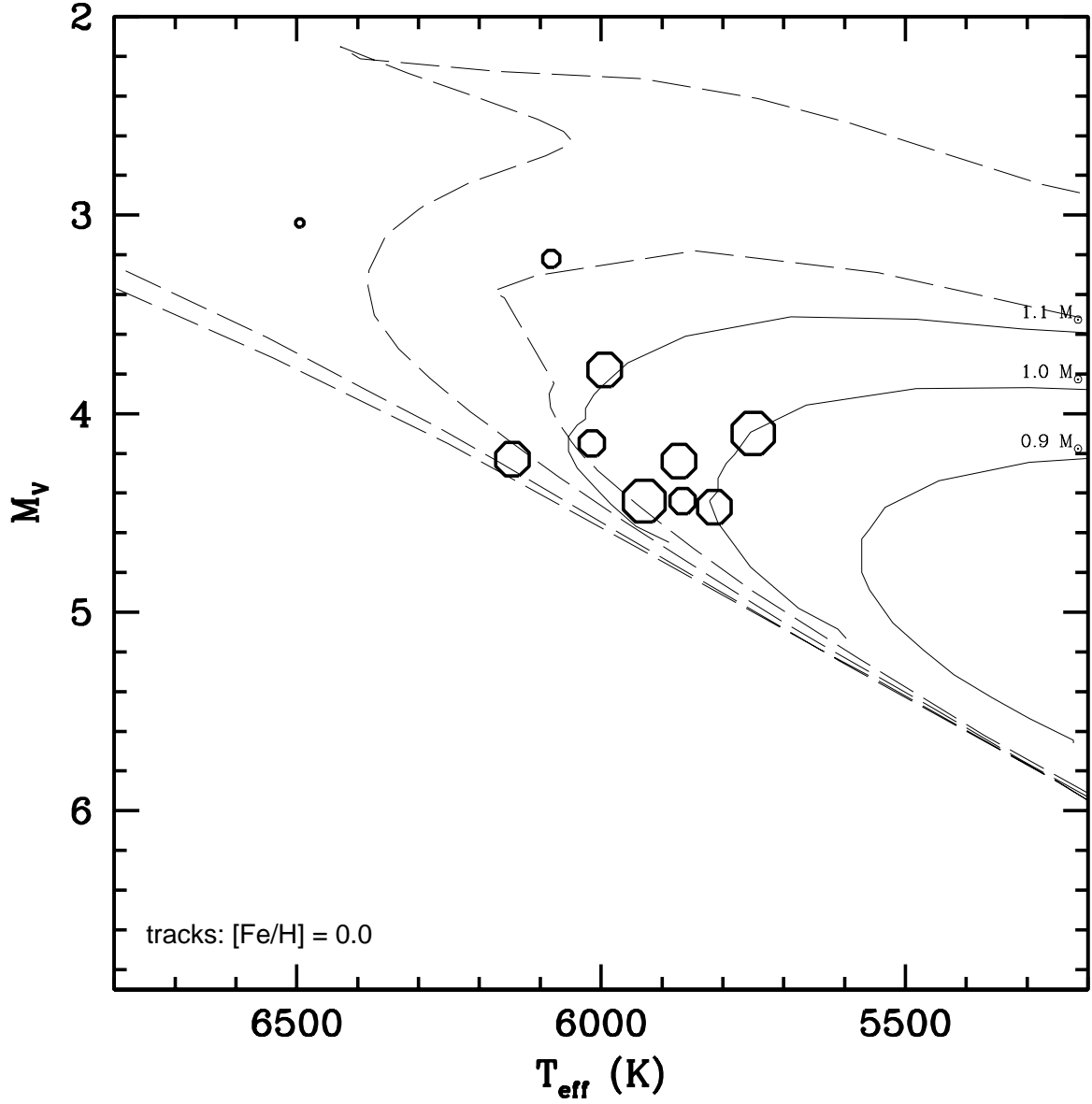}
\caption{HR diagram for the 10 stars with [Fe/H] $>$ $-$0.2.  Metallicity
dependent isochrones (dashed lines) for 100, 700, 2500, and 4500 Myr and
evolutionary tracks (solid lines) for [Fe/H] = 0 from Girardi et al. (2000)
are plotted.  The symbol size denotes the relative amounts of Be in each star,
with the largest symbols indicative of stars with A(Be) $>$ 1.25 dex and
subsequent smaller sizes corresponding to A(Be) bins of 0.25 dex.}
\end{figure}

\begin{figure}
\plotone{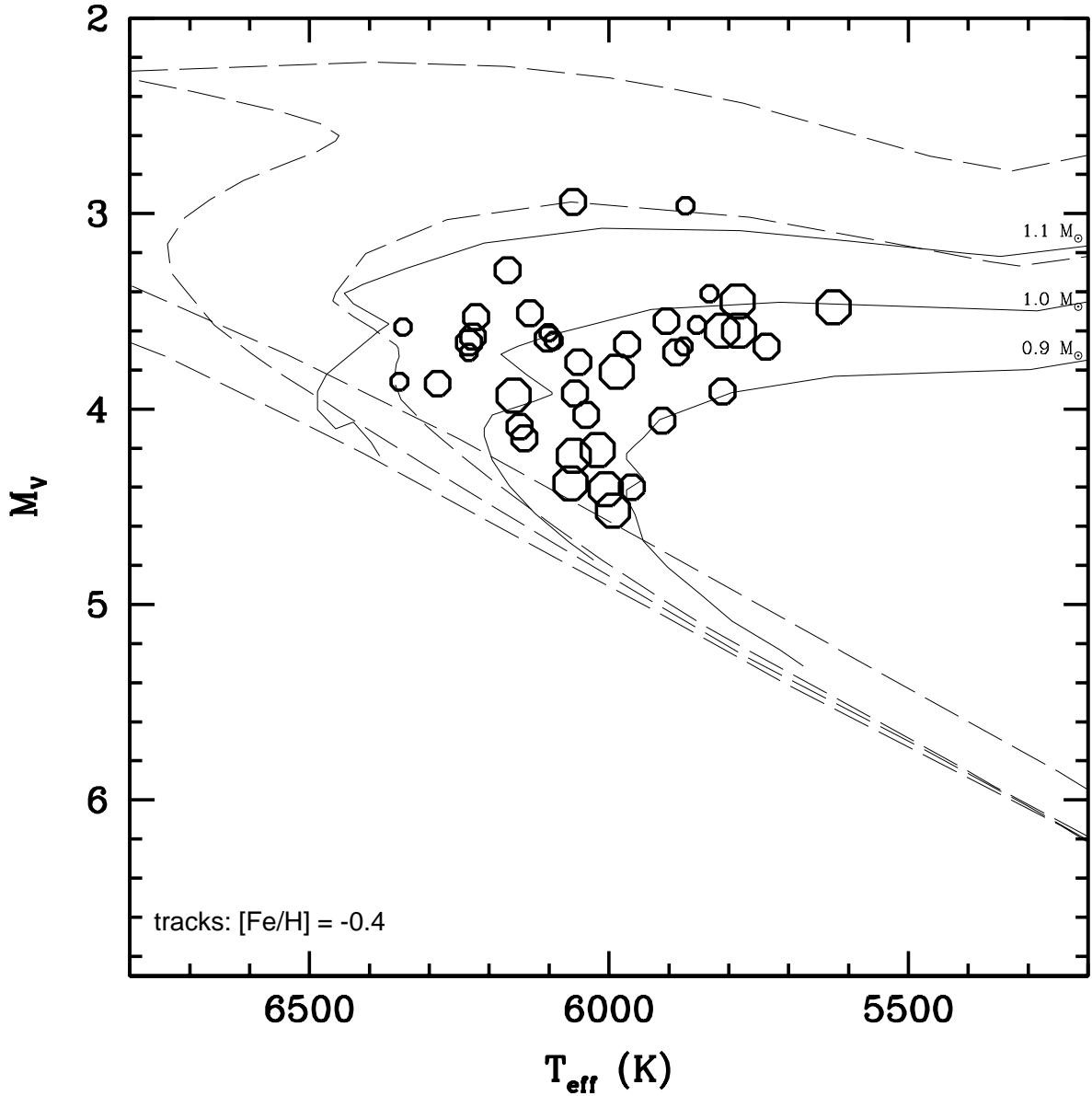}
\caption {HR diagram for stars with [Fe/H] $<$ $-$0.2.  Metallicity dependent
isochrones (dashed lines) for 100, 700, 2500, and 4500 Myr and evolutionary
tracks (solid lines) for [Fe/H] = $-$0.4 from Girardi et al.~(2000) are
plotted.  As in Figure 7, the symbol size indicates the relative amount of
Be.  In this case the largest symbol corresponds to 1.25 $\leq$ A(Be) $<$1.00
and the smallest indicates A(Be)$<$ 0.75.}
\end{figure}

\begin{figure}
\plotone{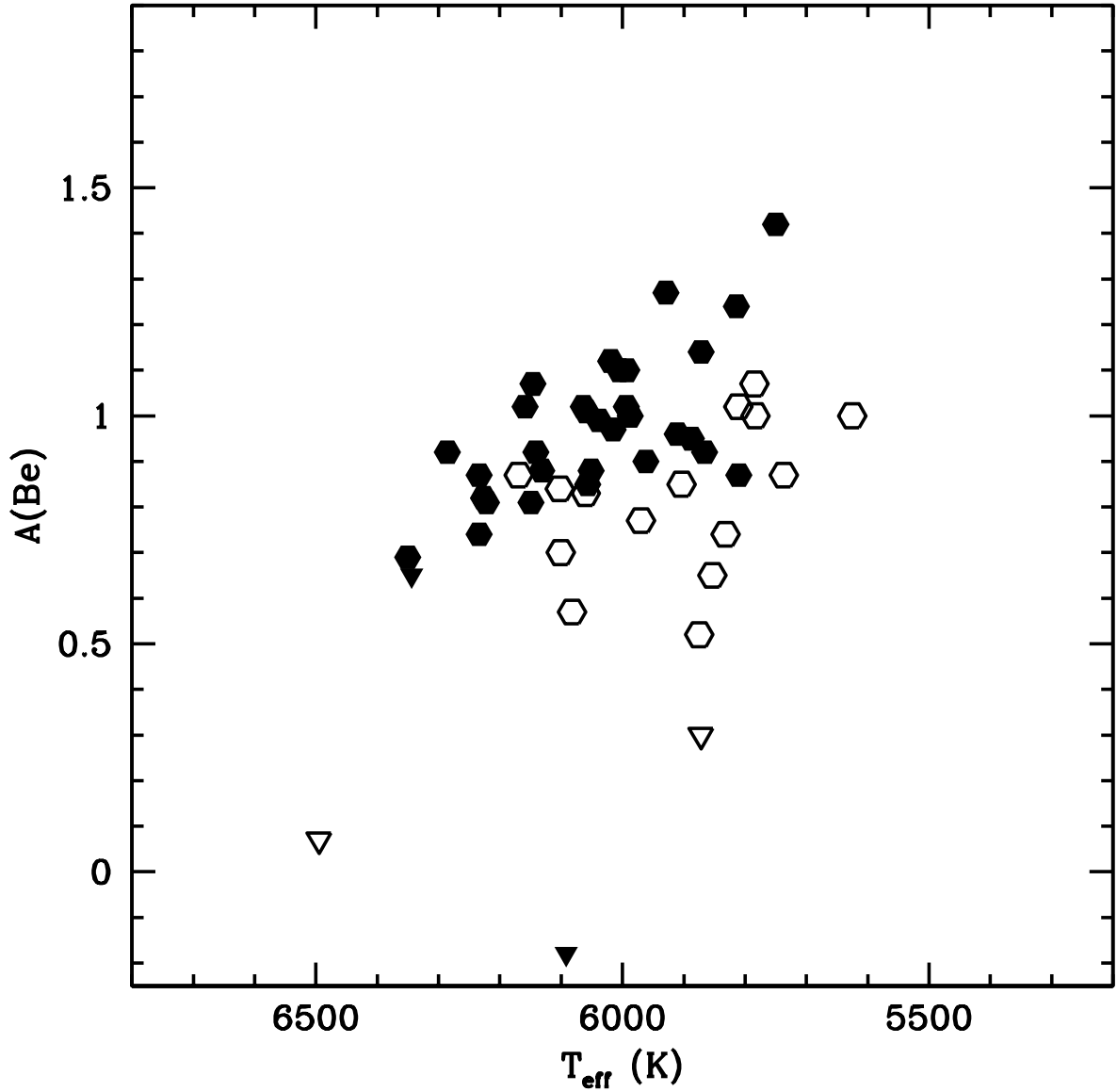}
\caption{Distribution of A(Be) with T$_{\rm eff}$.  An upper envelope seems to
be present showing an increase in A(Be) with decreasing temperature.  The
dwarf stars -- log g $>$ 4.1 -- are filled symbols while the subgiants are
open symbols.  The triangles are upper limits.  The subgiants show greater
depletion of Be at a given temperature.  This could indicate light element
dilution caused by the expansion of convection zone.}
\end{figure}

\begin{figure}
\plotone{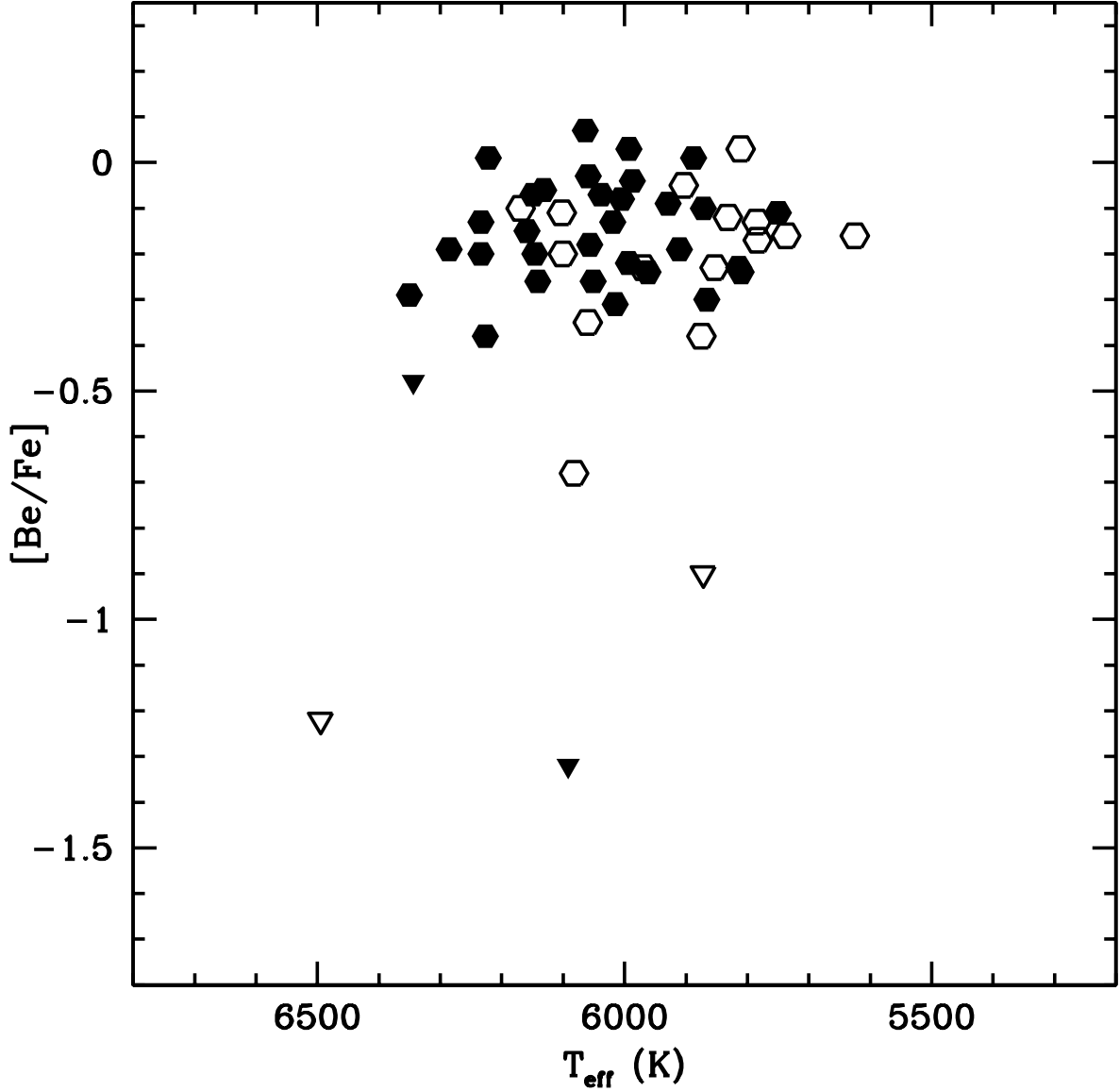}
\caption{Distribution of [Be/Fe] with T$_{\rm eff}$.  The dwarfs are filled
symbols and the subgiants are open symbols.  The upper envelope for detection
is not present when the Be abundance is normalized to the Fe abundance.  The
values of [Be/Fe] for dwarfs and subgiants appear to be similar.}
\end{figure}

\begin{figure}
\plotone{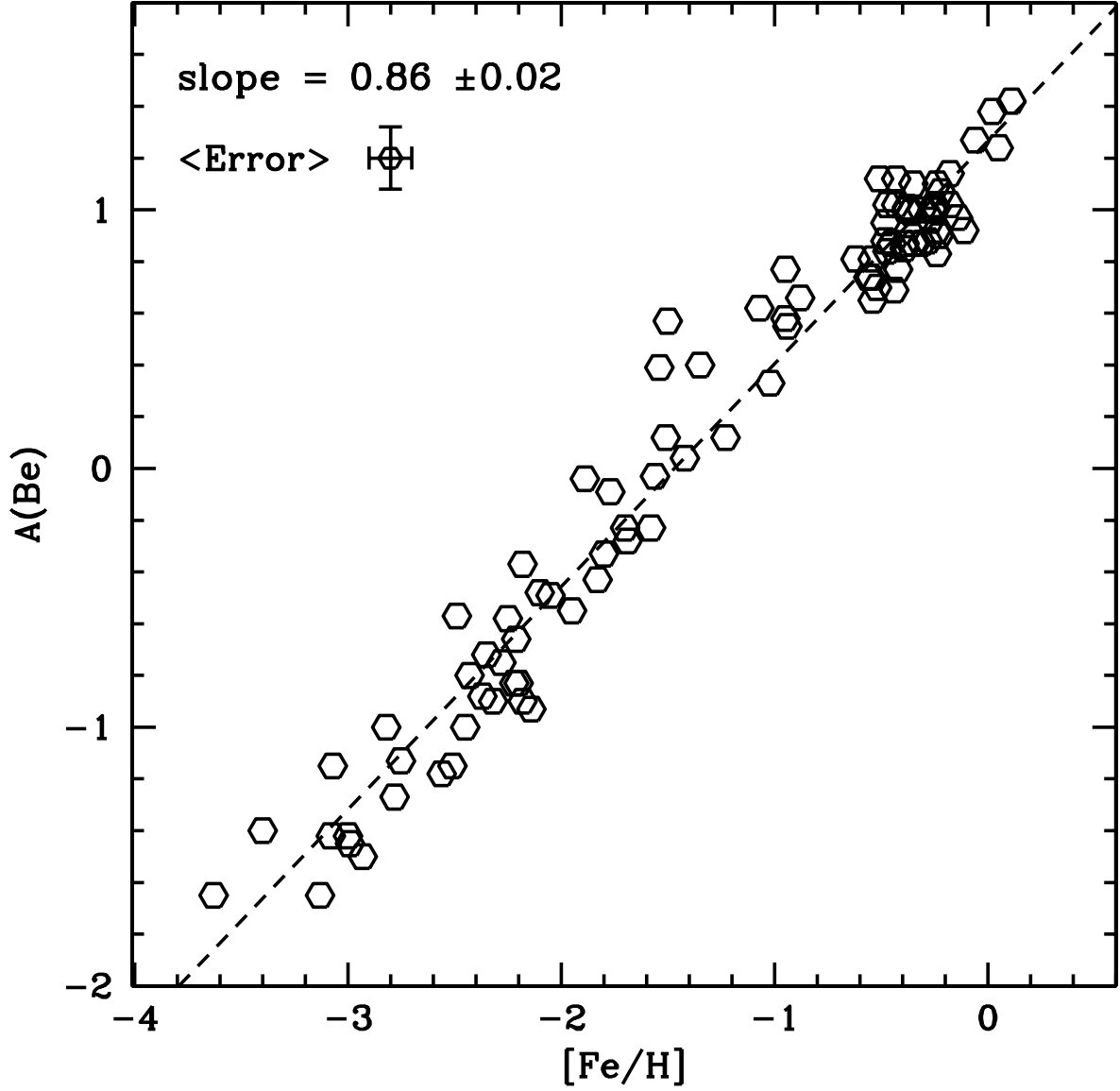}
\caption{Be abundances from this study along with Boesgaard et al.~(2008).  As
[Fe/H] increases, A(Be) increases with a slope of 0.86 $\pm$ 0.02, shown as
the dashed line.}
\end{figure}

\begin{figure}
\plotone{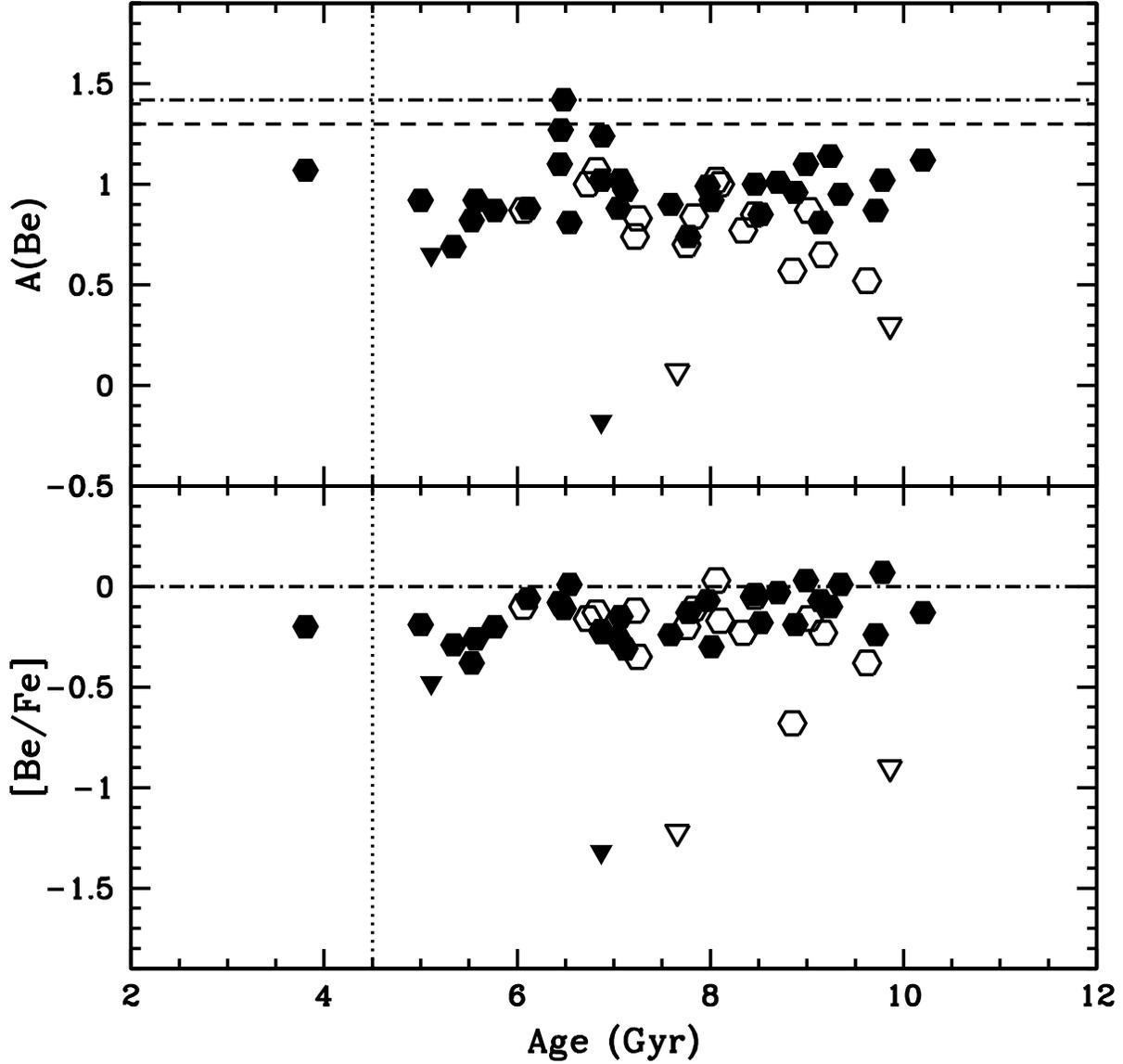}
\caption{Be plotted as a function of age.  The upper panel shows A(Be) while
the lower panel has the Be abundance normalized to Fe and to the solar values.
The meteoritic Be abundance, A(Be) = 1.42 is used for the solar system value.
The filled hexagons are the dwarf stars and the open hexagons are subgiants as
defined by log g $<$ 4.1.  The triangles are Be upper limits.  The vertical
dotted line represents the age of the Sun.  Only one star is younger than the
Sun.  In the upper panel the dashed line is placed at the approximate value of
the solar Be abundance (Boesgaard 2003) and the dashed-dotted line represents
the meteoritic Be abundance.  As mentioned in the text, most of the LR04 stars
are somewhat evolved off the zero age main-sequence and of the 157 stars in
the mass range 0.9 to 1.1 M$_{\odot}$ only 6 (or 4\%) are $<$ 5 Gyr in age.
Given that these stars are older than the sun and have lower Be abundances,
there is evidence for slow mixing during the main sequence stage of these
stars. }
\end{figure}

\begin{figure}
\plotone{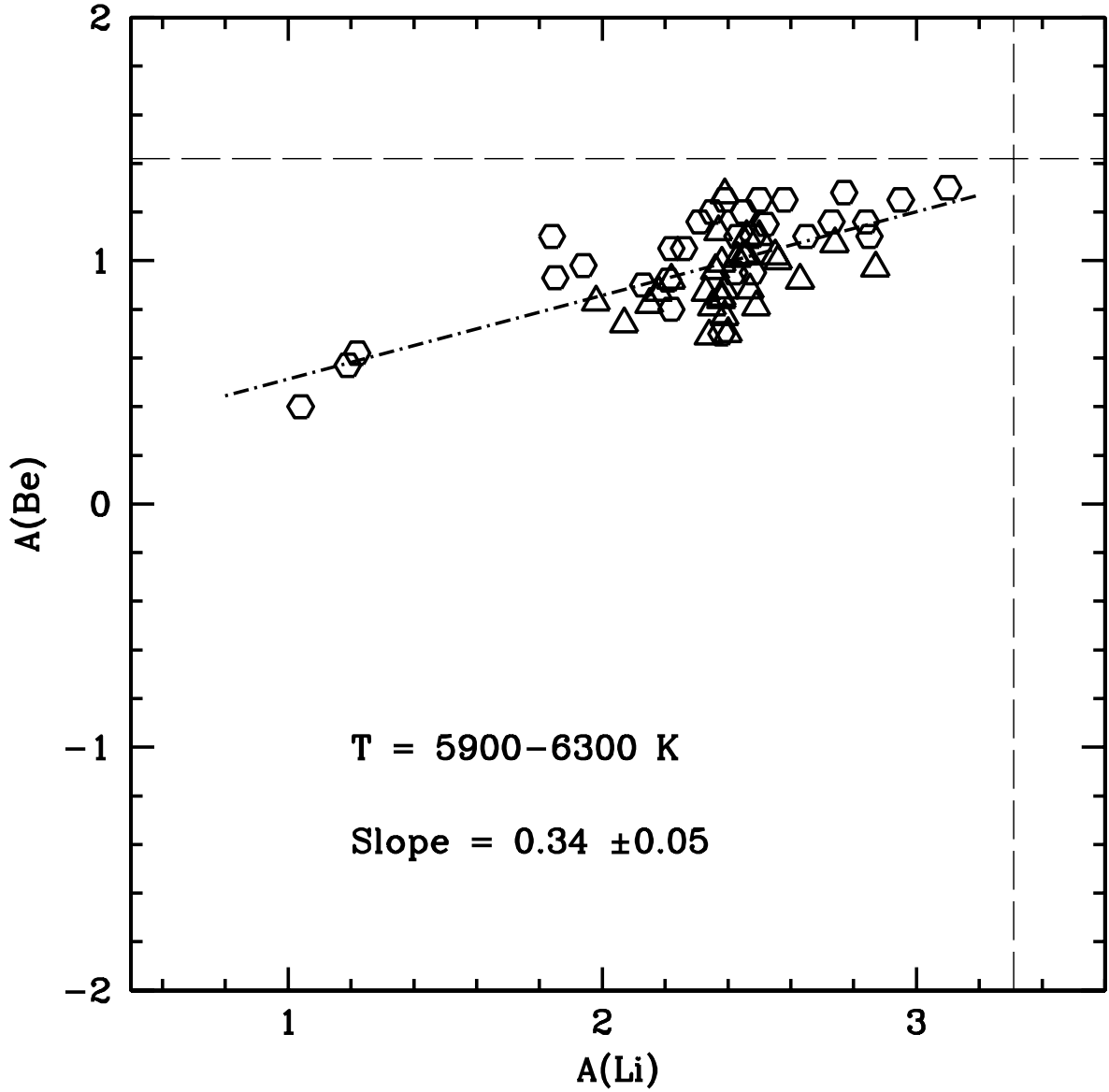}
\caption{A(Be) plotted against A(Li) with the B01 and B04ab data.  The
triangles represent the stars from this study in the T$_{\rm eff}$ range
5900-6300 K; the hexagons represent the sample from B01 and B04ab.  The one
solar-mass stars fit the relationship well.  The slope for all these stars is
+0.34 $\pm$0.05 which is consistent with the slope of 0.36 $\pm$ 0.05 given in
B04a.}
\end{figure}

\begin{figure}
\plotone{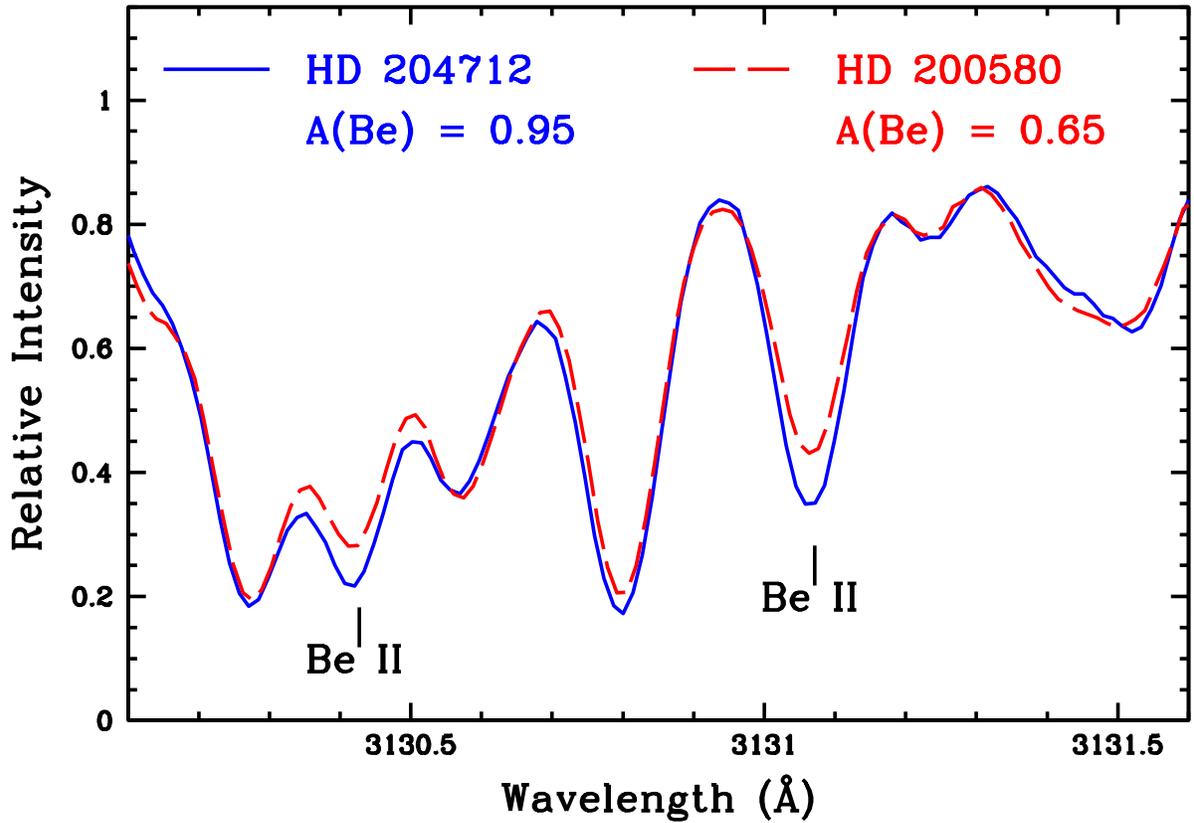}
\caption{The Be region of the observed spectra of HD 204712 and HD 200580.
The only difference in the two spectra is the strength of the two Be II
features.  The spectrum synthesis showed that these stars differ in their Be
abundances by a factor of 2.  They also differ in Li abundance by a factor of
two, with HD 204712 having both more Be and more Li.  The parameters of the
two stars are virtually identical with 204712/200580 having T$_{\rm eff}$ =
5888/5753; log g = 4.05/4.12; [Fe/H] = $-$0.48/$-$0.54; mass = 0.96/0.95
M$_{\odot}$; age = 9.35/9.17 Gyr.}
\end{figure}

\end{document}